\documentstyle[aps, preprint, eqsecnum]{revtex}

\renewcommand{\theequation}{\arabic{section}.\arabic{equation}}
\newcommand{\br}{{\bf r}}
\newcommand{\ba}{{\bf a}}
\newcommand{\bk}{{\bf k}}
\newcommand{\bX}{{\bf X}}
\newcommand{\dd}{{\cal D}}
\newcommand{\da}{{\cal A}}
\newcommand{\db}{{\cal B}}
\newcommand{\by}{{\bf y}}
\newcommand{\talpha}{{\tilde{\alpha'}}}
\newcommand{\tPi}{{\tilde{\Pi'}}}
\newcommand{\tb}{{\tilde{b}}}
\newcommand{\bv}{{{\bf u}}}
\newcommand{\bx}{{{\bf x}}}
\newcommand{\bE}{{{\bf E}}}
\newcommand{\bB}{{{\bf B}}}
\newcommand{\bH}{{{\bf H}}}
\newcommand{\sss}{{\sqrt{1-\bv^2}}}
\newcommand{\hbv}{{{\hat{\bf u}}}}

\begin{document}
\draft

\preprint{SCU-TP-97-1004;
SNUTP-97-107; CU-TP-854}
%GAC%%hep-th/9708149}

\title{Dynamics  of BPS Dyons: Effective Field
Theory Approach}

\author{ Dongsu Bak\footnote{email address: 
dsbak@mach.scu.ac.kr}}
\address{ Department of Physics,
University of Seoul,
Seoul 130-743, Korea} 
\author{Choonkyu Lee
%\footnote{email address: 
%cklee@phya.snu.ac.kr}
}

\address{Department of Physics, Seoul National 
University, Seoul 151-742, Korea}
\author{and\\Kimyeong 
Lee
%\footnote{email address: klee@phys.columbia.edu}
}
\address{Department of Physics, Columbia University, 
New York, NY 10027, USA}

\date{\today}

\maketitle
\widetext

\begin{abstract}
Based on a detailed analysis of nonlinear field equations of the SU(2)
Yang-Mills-Higgs system, we obtain the effective field theory
describing low-energy interaction of BPS dyons and massless particles
({\it i.e.}, photons and Higgs particles). Our effective theory
manifests electromagnetic duality and spontaneously broken scale
symmetry, and reproduces the multimonopole moduli space dynamics of
Manton in a suitable limit. Also given is a generalization of our
approach to the case of BPS dyons in a gauge theory with an arbitrary
gauge group that is maximally broken to ${\rm U(1)^k}$.

\end{abstract}


\section{Introduction}

In certain spontaneously broken non-Abelian gauge theories we have
magnetic monopoles as solitonic particles (in addition to usual
elementary field quanta) and, since their initial discovery by 't
Hooft and Polyakov\cite{thooft} in 1974, much effort has been made to
clarify their physical role. Then, more recently, a number of exact
results have been obtained in a class of supersymmetric gauge theories
by exploiting the electromagnetic duality symmetry\cite{seiberg}.
Magnetic monopoles relevant in this supersymmetric gauge theories are
so-called Bogomol'nyi-Prasad-Sommerfield (BPS) monopoles\cite{prasad},
{\it i.e.}, magnetic-charge-carrying static solutions to the
Yang-Mills-Higgs field equations in the BPS limit of vanishing Higgs
potential. In the BPS limit, there is a Bogomol'nyi bound on the
static energy functional and remarkably we have degenerate {\it
static} multi-monopole solutions that saturate the bound.  Originally
this was a semiclassical result at most; but, in the supersymmetric
gauge theories, Witten and Olive\cite{olive} subsequently showed that
this result may continue to be valid even after quantum corrections
are included.

To study the duality and other issues, various authors discussed the
interaction of slowly moving BPS monopoles, mainly following the work
of Manton\cite{manton1}. The central point is that the moduli space of
(gauge inequivalent) static $N$-monopole solutions is
finite-dimensional and possesses a natural metric coming from the
kinetic energy terms of the Yang-Mill-Higgs Lagrangian. Manton
suggested that low energy dynamics of a given set of monopoles and
dyons may be approximated by geodesic motions on the moduli space. The
metric for the two-monopole moduli space was determined by Atiyah and
Hitchin\cite{atiyah} and has given information as regards the
classical and quantum scattering processes of monopoles. More
recently\cite{sen}, the knowledge on the metric has been used in
theories with extended supersymmetry to show the existence of some of
the dyonic states required by the electromagnetic duality conjecture
of Montonen and Olive\cite{montonen}.

While Manton's approach is believed to give a valid approximate
description, it deviates from the viewpoint of modern effective field
theory---it is {\it not} based on all relevant degrees of freedom at
low energy. Dynamical freedoms in Manton's approach are restricted to
collective coordinates of monopoles, but the freedoms associated with
photons ($\gamma$) and massless Higgs particles ($\varphi$) are also
relevant at low energy.  We hope to remedy this in this article.
%Derivation of a simple effective theory involving 
%electromagnetic and Higgs fields together with 
%monopole collective coordinates
%is the main theme of this article. 
Instead of looking into the dynamics of collective coordinates of {\it
all} monopoles (this is Manton's moduli-space approximation), we will
here obtain our effective field theory by studying how the collective
coordinates of a {\it single} monopole/dyon get involved dynamically
with soft electromagnetic and Higgs field excitations in the vicinity
of the monopole/dyon.  This effective theory can describe the
low-energy interaction of monopoles with on-shell photons and Higgs
particles, and in the appropriate limit produces the result of Manton
as well.  (Note that, in our approach, monopoles/dyons interact
through the intermediary of electromagnetic and Higgs fields filling
the space.)  Moreover, it has distinctive advantage that underlying
symmetries of the theory, the electromagnetic duality and
spontaneously broken scale invariance, are clearly borne out, making
our effective action unique. 

The basic idea of our approach can be captured by considering the
low-energy effective theory of massive vector particles in the BPS
limit of SU(2) Yang-Mills-Higgs model.  In the unitary gauge with the
Higgs fields aligned as $\phi^a(x)=\delta_{a3}(f+\varphi (x))$, the
latter model is described by the Lagrange 
density\footnote{We set $c=1$, and our metric convention is that with
signature ($-$\,$+$\,$+$\,$+$).}
\begin{eqnarray}
\label{laguni}
&&{\cal L}=-{1\over 4} F^{\mu\nu}F_{\mu\nu}-{1\over 2}
|( \dd^\mu W^\nu-\dd^\nu W^\mu) |^2
-{1\over 2} \partial_\mu \varphi \partial^\mu \varphi 
-e^2 (f+\varphi)^2{W^\mu}^\dagger W_\mu\nonumber\\
&&\ \ \ +{ie}F^{\mu\nu}{W_\mu}^\dagger W_\nu +{e^2\over 4}
({W_\mu}^\dagger W_\nu-{W_\nu}^\dagger W_\mu) 
( {W^\mu}^\dagger W^\nu-
{W^\nu}^\dagger W^\mu)
\end{eqnarray}
where $F_{\mu\nu}=\partial_\mu A_\nu -\partial_\nu A_\mu$ is the
electromagnetic field strength, and $D_\mu W_\nu \,\,(\equiv
\partial_\mu W_\nu +ie A_\mu W_\nu)$ the covariant derivative of
charged vector field. The Higgs scalar $\varphi$, which is massless in
the BPS limit, plays the role of dilaton. When the energy transfer
$\Delta E$ is much smaller than the W-boson mass $m_v=ef$, the above
theory may be substituted by an effective theory with the action
$S_{\rm eff}$, whose dynamical variables consist of the
positions ${\bf X}_n(t)$ of W-bosons and two massless fields $A_\mu$
and $\varphi$. Ignoring contact interactions of `heavy' W-fields and
also relatively short-ranged magnetic moment interaction from
(\ref{laguni}), this low-energy action $S_{\rm eff}$ is easily
identified, {\it viz.},
\begin{eqnarray}
\label{lageff}
S_{\rm eff}=\int d^4 x \{{1\over 4} 
F^{\mu\nu}F_{\mu\nu}-{1\over 2}F^{\mu\nu}
(\partial_\mu A_\nu -\partial_\nu A\mu)
-{1\over 2} \partial_\mu \varphi \partial^\mu \varphi\} 
+\int dt L_{\rm eff}
\end{eqnarray}  
with $L_{\rm eff}$ given by
\begin{eqnarray}
\label{alageff}
L_{\rm eff}\!=
\!\sum_{n=1}^N \biggl\{\!-\!\left(m_v \!+\! g_s 
\varphi(\bX_n,t)\right)
\sqrt{1\!-\!{\dot \bX}_n^2}\!-
\!q_n [A^0(\bX_n,t)\!-\!{\dot\bX}_n(t)\cdot {\bf A}
(\bX_n,t)]\biggr\},
\end{eqnarray}  
where $q_n=\pm e$ and $g_s= {m_v\over f}= e>0$, denoting the electric
and dilaton charges of the W-particle, respectively. While we are
eventually interested in the low energy dynamics, it is also usuful to
keep the full relativistic kinetic terms for particles and solitons.
We remark that aside from the electromagnetic gauge invariance, this
effective theory also inherits from the original theory the
spontaneously broken scale invariance, which is described by
\begin{eqnarray}
\label{gaugetr}
&&m_v +g_s \varphi'(x)={1\over \lambda}
[m_v + g_s\varphi(x/\lambda)],\nonumber\\
&&A'_\mu(x)={1\over \lambda} A_\mu(x/\lambda), \ \ \ 
\bX'_n(t)=\lambda \bX_n(t/\lambda),
\end{eqnarray}  
where $\lambda$ is a real number.

{}From (\ref{alageff}) we see that the low-energy dynamics of
W-particles are governed by the force law (here, ${\bf V}_n\equiv
{d\over dt}\bX_n$)\footnote{As the force law
for the n-th W-particle, ${\bf E}$, ${\bf B}$ and ${\bf H}(=-\nabla
\varphi)$ appearing here may be allowed to include only contributions
which are really external to the very W-particle.}
\begin{eqnarray}
\label{weleq}
{d\over dt}\left[\left\{m_v\! +\!g_s \varphi(\bX_n,t)\right\}
{{\bf V}_n\over 
\sqrt{1\!-\!{\bf V}_n^2}}\right]=
q_n {\bf E}(\bX_n,t)\!+\! q_n{\bf V}_n\times 
{\bf B}(\bX_n, t)\!+\!
g_s {\bf H}(\bX_n, t)\sqrt{1\!-\!{\bf V}_n^2},
\end{eqnarray}  
where we have introduced the Higgs field strength ${\bf H}(x)\equiv
-\nabla \varphi (x)$ together with the electric and magnetic fields
$({\bf E}, {\bf B})$.  When nonrelativistic kinematics is appropriate,
(\ref{weleq}) reduces to
\begin{eqnarray}
\label{wnoneq}
m_v{d^2\over dt^2}{\bX}_n=
q_n [{\bf E}(\bX_n,t)+ {\bf V}_n\times {\bf B}(\bX_n, t)]+
g_s {\bf H}(\bX_n, t),
\end{eqnarray}
and then, as was done in the classical 
electrodynamics\cite{jackson}, 
one may use this force law
with field equations satisfied by $A_\mu$ and $\varphi$
to discuss various low-energy processes. 
Associated with a uniformly accelerating
W-particle with acceleration $\ba$, for instance, 
%there will be radiation fields
the usual near-zone fields will be accompanied by the 
radiation fields
\begin{eqnarray} 
\label{wradf}
&&{\bf E}(\br,t)\sim  {q_n\over 4\pi} 
{{\bf R}\times ({\bf R}\times \ba)\over R^3}, 
\ \ \ \ \ \ \ \  
{\bf B}(\br,t)\sim  -{q_n\over 4\pi} {{\bf R}\times 
\ba\over R^2}\nonumber\\
&&{\bf H}(\br,t)\sim  {g_s\over 4\pi} { ({\bf R}\cdot 
\ba){\bf R}\over R^3}, 
\ \ \ \ \ \ \ \  
 H^0(\br,t)\sim  {g_s\over 4\pi} {{\bf R}\cdot 
\ba\over R^2}
\end{eqnarray}  
where ${\bf R}$ is the radial distance vector 
evaluated at the retarded
time. Also the low-energy laboratory cross sections 
for the $\gamma W$ and 
$\varphi W$ scatterings are easily calculated to be
\begin{eqnarray} 
\label{wcrossection}
&&\left({d\sigma\over d\Omega}\right)_{{\gamma W}, \, {\varphi W}
\rightarrow {\gamma W}}=
\left({e^2\over 4\pi m_v}\right)^2 \sin^2\theta,\ \nonumber\\
&& 
\left({d\sigma\over d\Omega}\right)_{{\gamma W},\, {\varphi W}
\rightarrow {\varphi W}}=
\left({e^2\over 4\pi m_v}\right)^2 \cos^2\theta
\end{eqnarray}
where $\theta$ is the angle between the direction of outgoing massless
particles and that of the incident massless fields. Here we have
neglected the spin of $W$ particles. We have also taken care of the
photon spin by averaging over the initial spin and summing over the
final spin.  Of course the same results may be gotten in the tree
approximation of the full theory.

The above effective theory may also be used to 
derive the effective Lagrangian
for a system of slowly moving W-particles. 
%
%
%For the purpose, let $L_{(n)}$ denote
%the Lagrangian for the n-th W-particle in the potentials 
%$A_\mu$ and $\varphi$
%produced by the other W-particles in the system. Then, 
%assuming nonrelativistic 
%kinematics with (\ref{alageff}), it is 
%possible to write
%\begin{eqnarray}
%\label{nlageff}
%L_{(n)}=
%{1\over 2}m_v {\dot \bX}^2-q_n [A^0_n(\bX_n(t),t)\!-\!
%{\dot\bX}_n(t)\cdot {\bf A}_n
%(\bX_n(t),t)]-g_s \varphi_n (\bX_n,t)
%\end{eqnarray}  
%with near-zone potentials (as given in the textbook by 
%Landau and 
%Lifshitz\cite{landau},)
%\begin{eqnarray}
%\label{potential}
%&&A_n^0(\bX_n,t)=\sum_{n\neq m} 
%{q_m\over 4\pi |\bX_n-\bX_m|},\nonumber\\
%&&{\bf A}_n(\bX_n,t)=\sum_{n\neq m} {q_m\over 4\pi } 
%{{\dot\bX}_m + 
%({\hat U}_{nm}\cdot {\dot\bX}_m){\hat U}_{nm}
%\over|\bX_n-\bX_m|}, \ 
%{\hat U}_{nm}= {\bX_n-\bX_m\over |\bX_n-\bX_m|}
%\nonumber\\
%&&\varphi_n^0(\bX_n,t)=\sum_{n\neq m} {-g_s\over 4\pi 
%|\bX_n-\bX_m|}.
%\end{eqnarray}  
%The full Lagrangian, with its $\bX_n$-dependent part 
%given by $L_{(n)}$,
%then followes immediately\footnote{Our
%effective Lagrangian will lose its validity if two oppositely
%charged particles approach each other too closely.}:
%
%
%
This effective particle lagrangian results once we eliminate
massless fields $A_\mu(x)$ and $\varphi(x)$ from the above 
effective
Lagrangian by using their field equations in the near-zone 
approximation. For details on this procedure, 
see Appendix A. Assuming
nonrelativistic kinematics for $W$-particles, we then find the 
slow-motion Lagrangian of the 
form\footnote{Our
effective Lagrangian will lose its validity if two oppositely
charged particles approach each other too closely.}
%\begin{eqnarray}
%\label{modlag}
%L&=&
%{1\over 2}\sum_{n} {\dot \bX}^2_n  + 
%\sum_{n>m} {g_s^2 (\dot\bX_n-\dot\bX_m)^2 \over 8\pi|\bX_n-\bX_m|} 
%\nonumber\\
%&+&\sum_{n>m} {g_s^2-q_nq_m\over 8\pi|\bX_n-\bX_m|}
%(\dot\bX_n\cdot\dot\bX_m+U_{nm}\cdot\dot\bX_n U_{nm}
%\cdot\dot\bX_m)
%+\sum_{n>m} {g_s^2-q_nq_m\over 8\pi|\bX_n-\bX_m|} 
%\end{eqnarray} 
%with $U_{nm}$ being ${\bX_n-\bX_m/|\bX_n-\bX_m|}$.
\begin{eqnarray}
\label{modlag}
L=
{1\over 2}\sum_{n,m} g^{(nm)}_{ij}
(\bX){\dot \bX}_n^i  {\dot \bX}_m^j 
+\sum_{n>m} {g_s^2-q_nq_m\over 4\pi|\bX_n-\bX_m|}
\end{eqnarray} 
with the inertia metric 
\begin{eqnarray}
 g^{(nm)}_{ij}(\bX)&=&m_v \delta_{nm}\delta_{ij}
-{g_s^2\over 4\pi} \left[ \delta_{nm}\left(
\sum_{k(\neq n)}{1\over |\bX_k-\bX_n|}\right)
-{1-\delta_{nm}\over |\bX_n-\bX_m|}\right]
\delta_{ij} \nonumber\\
&+&{q_nq_m-g_s^2\over 8\pi|\bX_n-\bX_m|}\left[\delta_{ij}+ 
{(X_n^i -X_m^i)(X_n^j -
X_m^j)\over |\bX_n-\bX_m|^2}\right](1-\delta_{nm}).
\label{metric}
\end{eqnarray} 
In the special case of equally charged W-particles only, the potential
terms in (\ref{modlag}) cancel since $q_nq_m=g^2_s=e^2$, and also the
last term of the inertia metric (\ref{metric}). 
%so the
%system is governed solely by the kinetic Lagrangian $L={1\over
%%2}\sum_{n,m} g^{(nm)}_{ij}(\bX){\dot \bX}_n^i {\dot \bX}_m^j $
%$L=
%{1\over 2}\sum_{n} {\dot \bX}^2_n  + 
%\sum_{n>m} {g_s^2 (\dot\bX_n-\dot\bX_m)^2 \over8\pi
%|\bX_n-\bX_m|}$; 
with the metric $g^{(nm)}_{ij}(\bX)=m_v \delta_{nm}
\delta_{ij}
-{g_s^2\over 4\pi} \{ \delta_{nm}(\sum_{k(\neq n)}
{1\over |\bX_k-\bX_n|})
-{1-\delta_{nm}\over |\bX_n-\bX_m|} 
\}\delta_{ij}$;
%trajectories are now given by geodesics in the space 
%with this inertia metric.
One may discuss, for instance, 
low-energy scattering of two W-particles on 
the basis of
this effective Lagrangian.

In this paper we shall make a systematic study of the field equations
of the Yang-Mill-Higgs system to establish the low-energy effective
theory involving BPS monopoles/dyons. This will be much harder to
analyze than the case of the $W$-particles, for we here have to
confront the problems associated with {\it nonlinear} nature of the
given field equations.
%%%%straight. 
In the next section, static BPS dyon solutions are reviewed. Then, in
Sec.III, the force law analogous to (\ref{weleq}) will be derived for
a BPS dyon, and so are the appropriate generalizations of the results
(\ref{wradf}) and (\ref{wcrossection}) when BPS dyons, rather than
W-particles, are involved. Two of us have considered some part of these
problems earlier\cite{bak1,bak2}, but they did not encompass all the
relevant processes (especially those involving massless Higgs
particles). In Sec.IV, we then formulate the effective field theory
involving the dyon positions and two massless fields mentioned above,
in such a way that the results of Sec.III are fully accommodated. The
resulting theory assumes the form corresponding to a duality-invariant
generalization of the action (\ref{laguni}). It is conceivable that
our effective theory may have validity beyond tree level in the
context of appropriately supersymmetrized models. Also, for a system
of slowly moving BPS dyons (of the same sign), we obtain the effective
Lagrangian analogous to (\ref{modlag}) by the same procedure as above
and show that it is closely related to Manton's moduli-space dynamics
for well-separated monopoles. In Sec.V we discuss similar issues for
BPS dyons in a gauge theory with an arbitrary gauge group that is
maximally broken to $U(1)^k$. Here the appropriate monopole moduli
space was recently obtained in Ref. \cite{klee}.
%and we compare
%our effective field theory approach with this. 
Section VI is devoted
to the summary and discussion of our work. 

We have included brief reviews of some relevant materials to make our
paper reasonably self-contained. Presumably, various ideas developed
in this work were previously anticipated by Manton\cite{samols} and
others\cite{klee,harvey}, who presented a simple derivation of the
moduli-space metric for well-separated monopoles on the basis of
closely related ideas. But, up to our knowledge, the full story as
presented here  did not appear before.  In any case our work might be
viewed as the first first-principle derivation of the effective field
theory for the BPS monopoles and massless fields, in the sense that it
has been extracted through a detailed study of time-dependent dynamics
as implied by nonlinear field equations of the system.

\section{Static BPS Dyon Solutions in SU(2) Gauge Theory}

We shall here recall the basic construct of the BPS dyon solution in
an SU(2) gauge theory spontaneously broken to U(1).  For this
discussion it is better not to work in the unitary gauge. The
Lagrangian density is $(a = 1,2,3)$
\begin{equation}
\label{lag}
{\cal L} =-{1\over 4} G_a^{\mu\nu}G^a_{\mu\nu}-
{1\over 2} (D_\mu\phi)_a (D^\mu\phi)_a 
\end{equation}
where
\begin{eqnarray}
&&G_a^{\mu\nu} =\partial^\mu A^\nu_a-\partial^\nu A^\mu_a
+e\epsilon_{abc}A^\mu_b A^\nu_c,\\
&&(D_\mu \phi)_a=\partial_\mu\phi_a+e\epsilon_{abc} 
A_{\mu}^b\phi^c.
\end{eqnarray}
The field equations read
\begin{eqnarray}
\label{fieldeq}
&&(D_\mu G^{\mu\nu})_a =-e\epsilon_{abc}(D^\nu\phi)^b \phi^c,\\
\label{pfieldeq}
&&(D_\mu D^\mu \phi)_a=0 .
\end{eqnarray}
Without any nontrivial Higgs potential in the Lagrangian 
density,
this is a classically scale-invariant system. For this system,
spontaneous symmetry breaking is achieved by demanding 
the asymptotic boundary condition 
\begin{equation}
\label{bc1}
|\phi|=\sqrt{\phi_a \phi_a} \rightarrow f >0,\ \ \ 
{\rm as}\  r \rightarrow \infty .
\end{equation}
The unbroken U(1) will be identified with the 
electromagnetic gauge group
below. 

The above system admits static soliton solutions 
in the form 
of magnetic monopoles
(or, more generally, dyons), the stability of 
which is derived
from the topological argument. They
will carry some nonzero charges with respect 
to long-ranged fields.
To be explicit, we may define the electric 
and magnetic charges by
\begin{eqnarray}
\label{charge}
q =\oint_{{r= \infty}} dS_i \hat{\phi}^a E^a_i,\ \ \ 
g =\oint_{r= \infty} dS_i \hat{\phi}^a B^a_i
\end{eqnarray} 
with $E_i^a\equiv G^{0i}_a$, $B_i^a= {1\over 2}
\epsilon_{ijk} G^{jk}_a$ 
and $\hat{\phi}^a=\phi^a/\sqrt{\phi_a \phi_a}$, and the dilaton 
charge\footnote{This name is 
due to J.A. 
Harvey\cite{harvey}, who also emphasized the role of 
a Higgs scalar as a dilaton.} by
\begin{eqnarray}
\label{dilaton}
g_s =\oint_{r= \infty} dS_i \partial_i |\phi| 
=\oint_{r= \infty} dS_i \hat{\phi}^a (D_i\phi)^a
\end{eqnarray} 
Then we have $g={4\pi n /e}\, (n \in {\bf Z})$ for a 
topological reason while
$q$ may take on classically any continuous value. 
Also, $g_s$ is 
nothing but the mass of a static localized soliton up to a 
factor, {\it viz.},
\begin{equation}
\label{dilcharge}
g_s ={M/ f}
\end{equation}
with 
\begin{eqnarray}
\label{mass}
M \equiv \int d^3 r T^{00}=\int d^3r {1\over 2}
\left\{ E^a_i E^a_i\!+\! B^a_i B^a_i
\!+\!(D_0\phi)^a (D_0\phi)^a\! +\!(D_i\phi)^a 
(D_i\phi)^a \right\},
\end{eqnarray}
where $T^{00}$ denotes the 00-component of the 
stress energy tensor
\begin{equation}
\label{tensor}
T^{\mu\nu}=G_a^{\mu\lambda}G^{\nu}_{a\,\,\lambda}+
(D^\mu\phi)_a (D^\nu\phi)_a+\eta^{\mu\nu}{\cal L}. 
\end{equation}

The result (\ref{dilcharge}), which seems to be not very well-known,
can be proved as follows. Consider the so-called improved stress
energy tensor~\cite{callan}
\begin{equation}
\label{imtensor}
{\tilde{T}}^{\mu\nu}=T^{\mu\nu}+{1\over 6} 
(\eta^{\mu\nu}\partial^2-
\partial^\mu\partial^{\nu})|\phi|^2, 
\end{equation}
which is also conserved and satisfies at the same time the property of
being traceless, after using the field equations. Then, for any static
solution,
\begin{eqnarray}
\label{mass1}
\int d^3 r {\tilde{T}}^{00}&=&M-{1\over 6}\int d^3 
r\nabla^2 |\phi|^2\nonumber\\
&=&M-{1\over 3}f g_s, 
\end{eqnarray}
using (\ref{mass}) and the asymptotic behavior $|\phi|\sim f-{g_s\over
4\pi r}$.  On the other hand, since the traceless tensor
${\tilde{T}}^{\mu\nu}$ is also divergenceless, we have
\begin{eqnarray}
\label{mass2}
\int d^3 r {\tilde{T}}^{00}&=&\int d^3 r 
{\tilde{T}}^{ii}=
\int d^3r\partial_i({\tilde{T}}^{ij}x^j)\nonumber\\
&=& \oint_{r=\infty}dS_i {x^j\over 6}
(\delta_{ij}\nabla^2 -\partial^i\partial^j)
|\phi|^2= {2\over 3}f g_s
\end{eqnarray}
The relation (\ref{dilcharge}) follows immediately from (\ref{mass1})
and (\ref{mass2})\footnote{If the new tensor
${\tilde{T}}^{\mu\nu}$ were used to define the soliton mass, one would
end up with the mass value $2M/3$.  but we adhere to our definition
(\ref{mass}) for the soliton mass since this mass also enters the
equation of motion for a soliton (see the next two sections); the {\it
physical} mass is equal to $M$. }.

Based on (\ref{mass}), it is not difficult to show that the mass of
configurations with given $g$ and $q$ satisfies the following
inequality (called the Bogomol'nyi bound)\cite{prasad,coleman}
\begin{equation}
\label{bound}
M\geq f\sqrt{g^2+q^2}. 
\end{equation}
Moreover, to obtain static solutions to field equations
(\ref{fieldeq})and (\ref{pfieldeq}) with the lowest possible energy
$M= f\sqrt{g^2+q^2}$ for given $g=\mp 4\pi n/e$ ($n$: positive
integer) and $q=g\tan\beta$, it suffices to consider solutions to the
first-order Bogomol'nyi equations\cite{coleman}
\begin{equation}
\label{bogomol}
 B^a_i=\mp \cos\beta (D_i\phi)^a,\ \ E^a_i =
\mp \sin\beta (D_i \phi)^a, \ (D_0\phi)^a=0.
\end{equation}
These are equations relevant to BPS dyons and for 
$\beta= 0$ reduce to
the Bogomol'nyi equations for uncharged monopoles:
\begin{equation}
\label{bogomono}
 B^a_i=\mp (D_i\phi)^a, \ \ \  A_0^a=0.
\end{equation}
Actually all dyon solutions to (\ref{bogomol}), denoted as
(${\bar\phi}^a(\br; \beta), {\bar A}_i^a(\br;\beta), {\bar
A}_0^a(\br;\beta)$), can be obtained from the static monopole
solutions (${\bar\phi}^a(\br; \beta=0), {\bar A}_i^a(\br;\beta=0)$)
satisfying (\ref{bogomono}). This is achieved by the simple
substitution\cite{manton2}
\begin{eqnarray}
\label{dyon}
&& \bar\phi_a ({\bf r};\beta)=
\bar\phi_a (\br \cos\beta;0) ,\nonumber\\
&& \bar{A}^a_i ({\bf r};\beta)=\cos\beta 
\bar{A}^a_i (\br \cos\beta;0) ],\nonumber\\
&& \bar{A}^a_0 ({\bf r};\beta)=\mp\sin\beta 
\bar\phi^a (\br 
\cos\beta;0) .
\end{eqnarray}

The $n=\pm 1$ solutions to (\ref{bogomono}) are
well-known\cite{prasad}:
\begin{eqnarray}
\label{monopole}
&&\bar{A}^i_a( {\bf r};0) =
\epsilon_{aij}{\hat r_j\over er}(1-
{m_v r\over \sinh m_v r}),
\ \ \nonumber\\
&& \bar\phi_a ({\bf r};0)= \pm \hat 
r_a f (\coth m_v r -{1\over m_v r}).
\end{eqnarray}
These describe BPS one-(anti-)monopole 
solution, centered at the spatial 
origin, with $g=\mp 4\pi/e$ and mass 
$M=g_s f=4\pi f/e$. If the substitution
(\ref{dyon}) is made with these solutions, 
the results are the (classical)
BPS dyon solutions with $g=\mp 4\pi/e$, 
$q=\mp 4\pi \tan\beta/e$ and mass
$M=g_s f=4\pi f/(e\cos\beta)$. Being 
a Bogomol'nyi system, there are also static 
multi-monopole solutions satisfying 
(\ref{bogomono}). But, physically,
they may be viewed as representing 
configurations involving several of the 
fundamental $n=\pm 1$ monopoles described 
above. The latter 
interpretation is supported  by the 
observation that the dimension 
of the moduli space of 
solutions with $g=\mp 4\pi n/e$ is $4n$~\cite{weiberg}; 
this is precisely
the number one would expect for configurations 
of $n$ monopoles,
each of which is specified by three position 
coordinates and a U(1) phase
angle associated with dyonic excitations.

\section{Time-dependent Solutions Based on 
%Yang-Mills-Higgs 
Field 
Equations}

\subsection{Summary of our previous analyses}

We now turn to the study of low-energy dynamics 
involving BPS dyons, 
as dictated by the time-dependent field equations of 
the Yang-Mills-Higgs system.
Particularly important processes are those in which
a single BPS dyon interacts with electromagnetic and Higgs 
fields---they give most direct information on
the nature of effective interaction vertices involving
these freedoms. Some of these processes were previously 
analyzed by two of us\cite{bak1,bak2}, and in this subsection 
we shall recall the results obtained there.

%We shall here recall the results of our previous 
%analyses
%which are relevant to our discussion of low-energy 
%BPS dyon dynamics.

The first case concerns an accelerating BPS dyon in the
presence of a weak, uniform, electromagnetic field 
asymptotically\cite{bak2},
{\it viz.}, under the condition that
\begin{equation}
\label{asymp}
r\rightarrow\infty:\ \ \  {\phi^a\over |\phi|}B^a_i
\rightarrow ({\bf B}_0)_i,\ \ \ 
{\phi^a\over |\phi|}E^a_i\rightarrow ({\bf E}_0)_i.
\end{equation}
This generalizes the problem originally considered by
Manton~\cite{manton2} some time ago.  Due to the uniform asymptotic
fields present, the center of dyon is expected to undergo a constant
acceleration, namely, ${\bf X}(t)={1\over 2}\ba t^2$ (the acceleration
$\ba$ to be fixed posteriorly) in the reference frame with respect to
which the dyon has zero velocity at $t=0$. To find the appropriate
solution to the field equations (\ref{fieldeq}) and (\ref{pfieldeq}),
the following ansatz has been chosen in Ref. \cite{bak2}:
\begin{eqnarray} 
\label{ansatz}
&&\phi^a(\br, t) ={\tilde \phi}^a(\br';\beta),\nonumber\\ 
&&A_i^a(\br, t) =-t a_i {\bar A}^a_0({\br';\beta})+
{\tilde A}^a_i(\br';\beta), \nonumber\\
&&A_0^a(\br, t) =-t a_i {\bar A}^a_i({\br';\beta})+
{\tilde A}^a_0(\br';\beta)
\end{eqnarray}
with 
\begin{eqnarray} 
\label{ansa1}
&&{\tilde\phi}^a(\br';\beta) ={\bar \phi}^a(\br';\beta) + 
\Pi^a (\br';\beta),\ \   {\tilde A}^a_i(\br';\beta) =
{\bar A}^a_i(\br';\beta) + 
\alpha^a_i (\br';\beta), \nonumber\\
&& {\tilde A}^a_0(\br';\beta) =\mp \sin\beta\, 
{\bar\phi}^a(\br';\beta) + 
\alpha^a_0 (\br';\beta),
\end{eqnarray} 
where ${\br}'\equiv \br -{\bf X}(t)$, the functions (${\bar
\phi}^a(\br;\beta)$, ${\bar A}^a_\mu(\br;\beta)$) represent the static
dyon solution given by (\ref{dyon}) (with $g=\mp 4\pi/e$ and $q=
g\tan\beta$), and the yet-to-be-determined functions
($\Pi^a$,$\alpha_\mu^a$) are assumed to be $O(a)$ (or $O(B_0)$ or
$O(E_0)$).  Terms beyond $O(a)$ are ignored. Note that the functions
($\Pi^a$,$\alpha_\mu^a$) will account for the long-range
electromagnetic and Higgs fields as well as the field deformations
near the dyon core.

It then follows that the field equations (\ref{fieldeq}) are 
fulfilled if the functions ($\Pi^a$, $\alpha^a_\mu$) 
satisfy the equations
\begin{eqnarray}
\label{modbog}
&& {\tilde{B}}^a_i= \mp({\tilde{D}}_i+a_i)^{ab}
(\cos\beta {\tilde{\phi}}^b \pm \tan\beta \alpha^b_0), \\
\label{moda}
&&({\bar D}_i {\bar D}_i \alpha_0)^a=- e^2 \cos^2\beta 
\epsilon_{abc}\epsilon_{bdf}{\bar\phi}^c{\bar\phi}^f\alpha_0^d,
\end{eqnarray}
where ${\tilde{D}}^{ab}_i\equiv (D^{ab}_i)_{A^a\rightarrow 
{\tilde{A}}^a}$, 
 ${\tilde{G}}^{ji}_c\equiv (G^{ji}_c)_{A^a\rightarrow 
{\tilde{A}}^a}$, and 
the suppressed dependent variable is $\br'$. At the same time, 
the field strength $E^a_i$ to $O(a)$ is given by
\begin{equation}
\label{efield}
E^a_i(\br,t) =-t a_j {\bar G}^{ij}_a +({\tilde{D}}_i+a_i)^{ab}
{\tilde{A}}^b_0. 
\end{equation}
From these equations and the condition (\ref{asymp}), one 
finds that the 
acceleration $\ba$ should have the value given by
\begin{equation}
\label{lorentz}
M\ba=g {\bf B}_0 + q {\bf E}_0, \ \ \ (M= {4\pi 
f\over e\cos\beta})
\end{equation}
while the function $\alpha_0^a$ behaves asymptotically 
such as 
\begin{equation}
\label{alphaasymp}
r\rightarrow\infty:\ \ \  \alpha_0^a(\br';\beta) \rightarrow 
\mp \cos\beta (\sin\beta\, {\bf B}_0 -\cos\beta \,
{\bf E}_0)\cdot \br' {{\hat{r}}}^{a}.
\end{equation}
Note that (\ref{lorentz}) is the equation of motion 
in the dyon's
instantaneous rest frame, and the corresponding covariant 
generalization
\begin{equation}
\label{colorentz}
{d\over dt}\left({M{\bf V}\over 
\sqrt{1-{\bf V}^2}}\right)=
g ({\bf B}_0-{\bf V}\times {\bf E}_0) + q ({\bf E}_0+
{\bf V}\times {\bf B}_0),
 \ \ \ ({\bf V}\equiv {d\over dt}{\bf X})
\end{equation}
can also be secured by further considering the 
implication as the 
Lorentz boost of our ansatz (\ref{ansatz}) is performed.

The explicit, closed-form solution to (\ref{modbog}) 
and (\ref{moda}) 
has been given in Ref. \cite{bak2}. Because of its rather 
complicated structure, we shall here describe its 
characteristic features only. It is everywhere 
regular, with the 
fields near the dyon core ({\it i.e.}, at distance 
$d \sim {1/m_v}$ )
deformed suitably to match smoothly the long-range 
fields having 
simple physical interpretation. 
The physical contents of the long-range electromagnetic 
field is given in terms of $B^{\rm em}_i\equiv 
{\phi^a\over |\phi|} B^a_i$ 
and $E^{\rm em}_i\equiv {\phi^a\over |\phi|} E^a_i$, 
and that of the
long-range Higgs field by 
$H^\mu \equiv  -{\phi^a\over |\phi|} (D^\mu\phi)^a$.
These quantities are conveniently expressed using the 
{\it retarded distance}
${\bf R}=\br-{1\over 2}\ba t^2_{\rm ret}$ with $t_{\rm ret}$ 
determined 
(for a given $\br$ and $t$) through the implicit equation 
$t-t_{\rm ret}=|\br-{1\over 2}\ba t^2_{\rm ret}|\equiv R$. 
Explicitly, in the 
region $m_v r' \gg 1$,
\begin{eqnarray}
\label{radfa}
{\bf B}^{\rm em}(\br,t) &\sim& {\bf B}_0 +
{g\over 4\pi}{\hat{\bf R}\!-\!{\bf v}_{\rm ret}\over 
(1-\hat{{\bf R}}\cdot{\bf v}_{\rm ret})^3 R^2}-{q\over 4\pi}
{\hat{\bf R}\times {\bf v}_{\rm ret}\over R^2}
\nonumber\\
&+&\left\{ {g\over 4\pi}{\hat{\bf R}\times
(\hat{\bf R}\times \ba)\over 
R}-{q\over 4\pi}
{\hat{\bf R}\times {\bf a}\over R}\right\},\\
\label{radfb}
{\bf E}^{\rm em}(\br,t) &\sim& {\bf E}_0 +
{q\over 4\pi}{\hat{\bf R}\!-\!{\bf v}_{\rm ret}\over 
(1-\hat{{\bf R}}\cdot{\bf v}_{\rm ret})^3 R^2}+{g\over 4\pi}
{\hat{\bf R}\times {\bf v}_{\rm ret}\over R^2}\nonumber\\
&+&\left\{ {q\over 4\pi}{\hat{\bf R}\times
(\hat{\bf R}\times \ba)\over 
R}+{g\over 4\pi}
{\hat{\bf R}\times {\bf a}\over R}\right\},\\
\label{radfc}
{\bf H}(\br,t) &\sim& {g_s\over 4\pi}{\hat{\bf R}\!-\!
{\bf v}_{\rm ret}\over 
(1-\hat{{\bf R}}\cdot{\bf v}_{\rm ret})^3 R^2}
+\left\{ {g_s\over 4\pi}{(\hat{\bf R}\cdot \ba)
\hat{\bf R}\over 
R}\right\},\\
\label{radfd}
{H}^0(\br,t) &\sim& {g_s\over 4\pi}{\hat{\bf R}
\cdot{\bf v}_{\rm ret}
\over 
(1-\hat{{\bf R}}\cdot{\bf v}_{\rm ret})^3  R^2}
+\left\{ {g_s\over 4\pi}{(\hat{\bf R}\cdot \ba)\over 
R}\right\},
\end{eqnarray}
where $\ba$ is given by (\ref{lorentz}), 
${\bf v}_{\rm ret}\equiv \ba t_{\rm ret}$,
and $g_s=4\pi/(e\cos\beta)$ ({\it i.e.}, equals to 
the dilaton
charge of the dyon). Note that the expressions 
(\ref{radfa}) and (\ref{radfb})
are fully consistent with the electromagnetic fields 
of a pointlike 
dyon in motion and exhibit the manifest 
electromagnetic duality. 
[See (\ref{wradf}) for a comparison.]
This statement applies to both near-zone fields of 
$O(R^{-2})$ and 
radiation fields 
[the $O(R^{-1})$ terms marked by 
the curly brackets in (\ref{radfa})-(\ref{radfd})].  
%In (\ref{radfa})-(\ref{radfd}), terms inside the 
%%%%%%%%%curly brakets, all
%being $O(R^{-1})$, describe the radiation fields 
%%%%%%%%%accompanying 
%an accelerating dyon. {}From those, 
Now the radiation energy flux,
measured by the $0i$-component of the stress energy 
tensor, is given as 
\begin{eqnarray}
\label{radflux}
T^{0i}_{\rm rad}&=&G_a^{0k}G_a^{ik}+(D^0\phi)^a(D^i\phi)^a
=\epsilon^{ijk} E_j^{\rm em}B_k^{\rm em}+H^0H^i
\nonumber\\
&=& {g_s^2\over 16\pi^2 R^2} \left(|\ba\times
\hat{\bf R}|
+|\ba\cdot\hat{\bf R}|\right),
\end{eqnarray}
where we used the relation $g_s^2=g^2+q^2$.

In Ref.\cite{bak1}, an analogous perturbative scheme 
was used to study light 
scattering off a neutral BPS monopole in the 
long-wavelength limit. Here the incident electromagnetic 
wave is assumed to have magnetic field given as 
\begin{equation}
\label{emfield}
 {\bf B}^{\rm em}_{\rm in}= {\rm Re}\left[{iM\omega^2 
\over g}
{\bf a}\,
e^{i{\bf k}\cdot {\bf x}-i\omega t}\right] \ \ 
(\omega=|{\bf k}|, \, \bk\cdot\ba=0)
\end{equation}
where the frequency $\omega$ and the magnitude $a$ are 
taken to be sufficiently
small so that $\omega/m_v\ll 1$ and $\omega a\ll 1$. 
The center of the monopole
is then expected to undergo a non-relativistic
motion 
\begin{equation}
\label{motion}
{\bf X}(t) ={\rm Re}[-i\ba e^{-i\omega t}]
\end{equation}
(with the initial condition ${\bf X}(0)=0$). So, 
in this case, 
the solution
to the field equations (\ref{fieldeq}) may be sought on 
the basis of the 
ansatz
\begin{eqnarray} 
\label{comansatz}
&&\phi^a(\br, t) ={\bar \phi}^a(\br-{\bf X}) + 
{\rm Re}[\Pi^a (\br, t)],\ \ \ 
(\Pi^a(\br,t)={\tilde{\Pi}}^a (\br-{\bf X})
e^{-i\omega t}),\nonumber\\
&&A^a_i(\br, t) ={\bar A}^a_i(\br-{\bf X}) + 
{\rm Re}[\alpha^a_i (\br, t)],\ \ \ 
(\alpha^a_i(\br,t)=
{\tilde{\alpha}}^a_i (\br-{\bf X})e^{-i\omega t}),
\nonumber\\
&&A^a_0(\br, t) = 
{\rm Re}[{\tilde{\alpha}}^a_0 (\br-{\bf X})e^{-i\omega t}],
\end{eqnarray} 
where ${\bar A}^a_i(\br)$ and ${\bar\phi}^a(\br)$ 
represent the static BPS
monopole solution in (\ref{monopole}). The function 
(${\tilde{\Pi}}^a$, ${\tilde{\alpha}}^a_\mu$) are assumed 
to be $O(a\omega)$, and in 
the asymptotic region should account for the incident 
wave and outgoing radiations.

Using the ansatz (\ref{comansatz}) with field 
equations (\ref{fieldeq}) and 
(\ref{pfieldeq}) 
give rise to complicated differential equations for the 
functions 
(${\tilde{\Pi}}^a$, ${\tilde{\alpha}}^a_\mu$). But
as noted in Ref.\cite{bak1}, a great simplification is 
achieved
with the introduction of the functions $\beta^a_i(\br,t)$ 
by the equation
\begin{equation}
\label{defaa}
 G_a^{ij}(\br, t)= \mp\epsilon_{ijk}[(D_k\phi)^a(\br,t)+
\beta_k^a(\br,t)].
\end{equation} 
The field equations are fulfilled if $\beta^a_i$  
satisfy the equation
\begin{equation}
\label{helmholtz}
[({\bar D}_k{\bar D}_k+\omega^2) {\tilde\beta}_i]^a
+e^2 
\epsilon_{abc}\epsilon_{bde} {\tilde{\beta}}_i^d
{\bar\phi}^e{\bar\phi}^c=0,
\end{equation}
(here, ${\bar D}^{ac}_i\equiv \partial_i\delta_{ac}+
e\epsilon_{abc}{\bar A}^b_i(\br-{\bf X})$), and 
then the functions 
${\tilde\phi}^a$ and ${\tilde A}^a_i$ can be found using 
\begin{eqnarray} 
\label{perteq2}
&&{\tilde\Pi}^a={1\over \omega^2}\left[({\bar D}_i 
{\tilde\beta}_i)_a -ie\omega 
\epsilon_{abc}{\tilde\alpha}^b_0{\bar\phi}^c- 
i\omega^2 a_j \partial_j
{\bar\phi}^a\right],\\
&&{\tilde\alpha}^a_i={1\over \omega^2}
\left[\mp\epsilon_{ijk}(D_j 
{\tilde\beta}_k)^a
+e 
\epsilon_{abc}{\tilde\beta}^b_i
{\bar\phi}^c-i\omega ({\bar D}_i
{\tilde\alpha}_0)^a-i\omega^2 a_j \partial_j
{\bar A}^a_i\right].
\label{perteq33}
\end{eqnarray}  
So what remains nontrivial is to solve 
(\ref{helmholtz}). There
is no equation to fix the functions $\alpha^a_0$, 
but this just reflects 
arbitrariness in the choice of gauge.

The solution to (\ref{helmholtz}), found in 
Ref.\cite{bak1}, reads
\begin{equation}
\label{solutiona}
 {\tilde\beta}_{i}^a(\br')=\pm i\omega^2 a_i f \coth m_v r' 
e^{i {\bf k}\cdot \br'}{\hat{r'}}^a \mp i\omega^2 a_i
{e^{i\omega r'}\over er'}{\hat{r'}}^a
+ O(a\omega^3),
\end{equation}
where $\br'\equiv \br-{\bf X}$. Then, using 
this with (\ref{perteq2}) 
and (\ref{perteq33}) 
(with the gauge choice 
${\tilde\alpha}^b_0(\br')= \omega a_i 
{\bar A}^b_i(\br')$),
made for the consistency of our ansatz), the 
expressions for 
${\tilde\Pi}^a(\br')$ and ${\tilde\alpha}^a_i
(\br')$ follow. In this way,
long-range fields in the present process 
have been identified as 
\begin{eqnarray} 
&&{\bf B}^{\rm em}(\br,t)\sim  \mp
 i\omega^2{\bf a} f e^{i\bk\cdot\br -iwt}
\mp i\omega^2
({\hat\br}\times({\hat\br}\times \ba))
{e^{i\omega r -iwt}\over er},\nonumber\\
&&{\bf E}^{\rm em}(\br,t)\sim \pm i\omega^2 
({\bf \hat k}\times {\bf a})
 f e^{i\bk\cdot\br -iwt}\mp i\omega^2
({\hat\br}\times \ba){e^{i\omega r -iwt}
\over er},\nonumber\\
\label{asympa}
&&{\bf H}(\br,t)\sim i\omega^2 
{\ba\cdot {\hat\br}\over er} e^{i\omega r -iwt}{\hat\br},\\
&&H^0(\br,t)\sim i\omega^2{\ba\cdot 
{\hat\br}\over er}e^{i\omega r -iwt}\nonumber
\end{eqnarray} 
where only the real parts are relevant. 
Notice the appearance of 
outgoing spherical waves, describing 
electromagnetic and Higgs scalar
radiations. Based on these results, the related 
differential
cross sections are determined as
\begin{eqnarray} 
\label{crossection}
&&\left({d\sigma\over d\Omega}\right)_{{\rm em}}=
{\left(\omega^4/2e^2\right)|{\hat r}\times \ba|^2\over 
{1\over 2}\omega^4 f^2 a^2}=
\left({g^2\over 4\pi M}\right)^2 \sin^2\Theta,\\
\label{acrossection}
&& \left({d\sigma\over d\Omega}\right)_{{\rm Higgs}}=
{\left(\omega^4/2e^2\right)|\ba\cdot{\hat r}|^2\over 
{1\over 2}\omega^4 f^2 a^2}=
\left({g_s^2\over 4\pi M}\right)^2 \cos^2\Theta
\end{eqnarray}  
where $\Theta$ is the angle between ${\hat{\bf r}}$ 
(i.e. the observation direction) and the incident 
${\bf B}^{\rm em}$-field, and 
we used the relation $g^2=g_s^2=(4\pi/e)^2$ here.
Notice a close similarity between these results for 
a BPS monopole and
the corresponding formulas in (\ref{wcrossection}) 
for an electrically
charged $W$-particle.

\subsection{Accelerating dyon solution in weak uniform 
asymptotic fields}

A BPS dyon, having none-zero dilaton charge, will have a 
nontrivial
coupling to the Higgs field. To deduce the corresponding 
force law 
from the field equations in a convincing way, it is 
necessary to consider more
general, uniform, asymptotic
field than in (\ref{asymp}). In this section, 
we therefore suppose 
that there exists also a weak, uniform, Higgs 
field strength
asymptotically, {\it viz.}, 
\begin{equation}
\label{asympa1}
{\phi^a\over |\phi|}(D_i\phi)^a\rightarrow \ 
-({\bf H}_0)_i\ \ \ 
{\rm as}\ r\rightarrow\infty
\end{equation}
in addition to the electromagnetic field 
strengths $({\bf E}_0,{\bf B}_0)$
specified as in (\ref{asymp}). Of course, the 
imposition of (\ref{asympa1}) 
would make the asymptotic condition (\ref{bc1}), 
required for any field
configuration with finite total energy, obsolete. 
This is not a problem;
our interest here is in studying time-dependent 
flow of
energy from one spatial region to another, as 
predicted by field equations. 
For sufficiently small (${\bf E}_0$,${\bf B}_0$,
${\bf H}_0$), we may again
seek the appropriate perturbative solution to 
the field equations on the basis of
the ansatz given in (\ref{ansatz}) and 
(\ref{ansa1}). This will lead to 
(\ref{modbog}) and (\ref{moda}), and 
also to the relation (\ref{efield}) 
for $E^a_i$. But the solution of our 
present interest is, unlike that given
in Ref.\cite{bak2}, the one satisfying 
(\ref{modbog})-(\ref{moda}) 
for non-zero ${\bf H}_0$.

Our first task is to determine the dyon 
acceleration $\ba$ under 
this generalized asymptotic condition. For this purpose,
we assume the asymptotic form of the function 
$\alpha^a_0$ to be given as
\begin{equation}
\label{asyme}
r\rightarrow\infty:\ \ \  \alpha^a_0(\br';\beta)\rightarrow 
\cos\beta \,{\bf C}\cdot\br'\, {\hat{r'}}_a
\end{equation}
(${\bf C}$ is some constant vector), and then we have 
\begin{equation}
\label{asymee}
r\rightarrow\infty:\ \ \  {\phi^a\over |\phi|} 
({\tilde D}_i\alpha_0)^a 
\rightarrow \pm\cos\beta\, C_i.
\end{equation}
Now we use this information and the given asymptotic 
conditions
with (\ref{modbog}) and (\ref{efield}), to deduce 
two linear relations
involving ${\bf B}_0$, ${\bf E}_0$, ${\bf H}_0$, 
${\bf a}$, and ${\bf C}$.
Solving the latter for  ${\bf a}$, and ${\bf C}$, we 
immediately obtain
\begin{equation}
\label{motioneq}
\ba=\mp{1\over f}[ \cos\beta \,{\bf B}_0 +\sin\beta 
\,{\bf E}_0\mp {\bf H}_0]
\end{equation}
and
\begin{equation}
\label{alpha}
{\bf C}=\mp{1\over f}[ \sin\beta\, {\bf B}_0 -\cos\beta\, 
{\bf E}_0  ]
\end{equation}
Notice that ${\bf C}$ does not depend on ${\bf H}_0$. 
If (\ref{motioneq})
is rewritten using $g=\mp 4\pi/e$, $q=\mp 4\pi
\tan\beta /e$ and 
$g_s=4\pi/(e\cos\beta)=M/f$, it assumes the form 
\begin{equation}
\label{dylorentz}
M\ba= g {\bf B}_0 +q 
{\bf E}_0+g_s {\bf H}_0.
\end{equation}
This is the desired equation of motion for a dyon 
in its 
instantaneous rest frame. We here remark that, by 
considering the
 Lorentz boost of the above solution, (\ref{dylorentz}) 
may be 
generalized to the form (see the Appendix B)
\begin{equation}
\label{dycolo}
{d\over dt}\left({(M-g_s X_\mu H^\mu){\bf V}\over 
\sqrt{1-{\bf V}^2}}
\right)=
g ({\bf B}_0-{\bf V}\times {\bf E}_0) + q ({\bf E}_0+
{\bf V}\times {\bf B}_0)+g_s{\bf H}_0 \sqrt{1-{\bf V}^2}.
\end{equation}
This should be compared with the force law for a W-particle, 
given in (\ref{weleq}).

If the strengths of the asymptotic fields are such that
\begin{equation} 
\label{staticcon}
{\bf H}_0=\pm( \cos\beta\, {\bf B}_0 +\sin\beta\, 
{\bf E}_0 ),
\end{equation} 
we see from (\ref{motioneq}) that $\ba =0$, {\it i.e.}, 
the dyon does not
feel any force (at least to the first order in 
the applied fields).
In view of (\ref{ansatz}), the corresponding 
solution is necessarily static.
Here one has the special case where the applied 
fields are consistent with 
the original Bogomol'nyi equations (\ref{bogomol}). This 
happens if $\alpha_0^a=0$ (and hence ${\bf C}=0$) and
\begin{equation}
\label{dualstatic}
{\bf B}_0=\pm\cos\beta\, {\bf H}_0, \ \ \  
{\bf E}_0= \pm\sin\beta\, {\bf H}_0.
\end{equation} 
We are now talking about a static BPS dyon
 solution in the presence
of {\it self-dual uniform fields}. After some 
calculation we have found
that the appropriate static solution, for 
$\beta=0$ ({\it i.e.} 
the case of a neutral monopole) and to $O(H_0)$, 
is given by
\begin{eqnarray}
\label{mowfield}
 \phi_a ({\bf r})&=& \pm \hat r_a f (\coth m_v r 
-{1\over m_v r})
\pm {1\over 2}{\bf H}_0\cdot \!\br 
\,{\hat r}_a {m_v r\over \sinh^2 m_v r}\nonumber\\
 &\mp& {1\over 2}(({\bf H}_0)_a-{\bf H}_0\!\cdot 
{\hat\br}\, {\hat r}_a)
{ r\over \sinh m_v r}
\mp \coth m_v r \,{\bf H}_0\cdot \br\, {\hat r}_a\ ,\\ 
A^i_a( {\bf r}) &=&
\epsilon_{aij}{\hat r_j\over er}(1-{m_v r\over 
\sinh m_v r})+
{1\over 2}\epsilon_{aij}{\hat r}_j {\partial
\over\partial r}\left(
{ r^2\over \sinh m_v r}\right){\bf H}_0\cdot 
{\hat\br}\nonumber\\
&+&\epsilon_{aij}(({\bf H}_0)_j-{\bf H}_0\!\cdot 
{\hat\br}\, {\hat r}_j)
{ r\over 2\sinh m_v r}+{ r-r\cosh m_v r\over 2
\sinh m_v r}{\hat r}_a 
\epsilon_{ilm} {\hat r}_l({\bf H}_0)_m,\\
A^a_0(\br)&=&0.
\end{eqnarray}
The corresponding solution for $\beta \neq 0$ 
({\it i.e.}, the BPS dyon case)
then also follows once the trick in (\ref{dyon}) 
is used\footnote{While (\ref{mowfield}) is only an 
approximate solution 
({\it i.e.}, valid to $O(H_0)$) of the Bogomol'nyi 
equation, we remark 
that its $m_v \rightarrow \infty$ limit, namely, 
\begin{eqnarray}
 \phi_a ({\bf r})&=& \pm {\hat r}_a (-{\bf H}_0 
\cdot \br +
f-{1\over er}),\nonumber\\
 A^i_a( {\bf r}) &=&
\epsilon_{aij}{{\hat r}_j\over er}-
{1\over 2}{\hat r}_a \epsilon_{ilm} x_l({\bf H}_0)_m,
\nonumber
\end{eqnarray}
corresponds to an exact, but singular, 
solution of $B^a_i=\mp (D_i\phi)^a$.}.

We now consider the solution to (\ref{modbog}) 
and (\ref{moda}) when $\ba$
is nonzero; this will lead to a time-dependent 
solution, accompanied by 
suitable radiation fields. Following 
Ref.\cite{bak2}, we 
introduce rescaled quantities 
\begin{equation}
\label{rescale}
{\bf y} ={\br'}\cos\beta, \ \ \ \da^a_i(\by)=
{1\over \cos\beta} {\tilde A}^a_i (\br'={\by\over 
\cos\beta};\beta)
\end{equation}
and recast (\ref{modbog}) and (\ref{moda}) as
\begin{eqnarray}
\label{pmodbog}
&& {\db}^a_i= \mp(\dd^{(y)}_i+{a_i\over \cos\beta})^{ab}
({\tilde{\phi}}^b \pm {\sin\beta\over 
\cos^2\beta} \alpha^b_0), \\
\label{pmoda}
&&({\bar D}^{(y)}_i {\bar D}^{(y)}_i 
\alpha_0)^a=- e^2 
\epsilon_{abc}\epsilon_{bdf}
{\bar\phi}^c(\by;\beta=0){\bar\phi}^f
(\by;\beta=0)\alpha_0^d,
\end{eqnarray}
where ${\bar{D}}^{(y^i)ac}_i\equiv 
{\partial\over \partial y^i}\delta_{ac}+
e\epsilon_{abc} {\bar A}^b_i(\by;\beta=0)$, 
${\dd}^{(y)ab}_i\equiv 
{\partial\over \partial y}\delta_{ac}+
e\epsilon_{abc} {\da}^b_i(\by)$, and 
$\db^a_i(\by)$ denotes 
the magnetic field strength obtained from the 
vector potential
$\da^a_i(\by)$. The solution to (\ref{pmoda}) which 
fulfills 
the condition  (\ref{asyme}) is given by 
\begin{equation}
\label{solalpha}
\alpha^a_0= \coth m_v y ({\bf C}\cdot \by) {\hat y}_a +
{y\over \sinh m_v y} [({\bf C})_a- ({\bf C}\cdot \hat\by)
{\hat y}_a],
\end{equation} 
where $y\equiv |\by|=r'\cos\beta$ and the 
vector ${\bf C}$ is given by 
(\ref{alpha}). We have also solution to 
(\ref{pmodbog}) expressed as
\begin{eqnarray}
\label{dyona}
 {\tilde\phi}_a &=& \pm {\hat y}_a \left[f 
(\coth m_v y -{1\over m_v y} )
 +{{\hat\by}\cdot \ba\over 2e\cos\beta}
(1-{m_v y\over \sinh m_v y})\right]
\pm {a_af y\over 2\cos\beta \sinh m_v y}\nonumber\\
 &\mp&{\sin\beta\over \cos^2\beta}\alpha^a_0
\mp {\hat y}_a{\partial\over\partial y}
(y\coth m_v y\, V)\mp
\left({\partial\over\partial y^a}-{\hat y}_a{\partial
\over\partial y} \right) 
\left( {y\over \sinh m_v y}V\right),\\
\label{dyonb} 
\da^i_a &=&
\epsilon_{aij}{\hat y_j\over ey}(1-{m_v y\over 
\sinh m_v y})+
 {fy\over 2\cos\beta}\left(\coth m_v y -
{1\over m_v y}\right) 
\epsilon_{ijk} {\hat y}_j a_k {\hat y}_a \nonumber\\
&+&
\epsilon_{aij}{\partial\over\partial y^j}\left(
{ y\over 2\sinh m_v y}V\right)+
(1-\cosh m_v y){\hat y}_a 
\epsilon_{ilm} {\hat y}_l {\partial\over\partial y^m}  
\left( {y\over \sinh m_v y}V\right),
\end{eqnarray}   
where the function $V$, which is adjustable, must 
satisfy the Laplace equation $\nabla^2 V=0$. All 
asymptotic 
boundary conditions, including (\ref{asympa1}), are 
satisfied if we here choose
\begin{equation}
\label{laplace}
V =-{1\over 2\cos^2\beta}(\sin\beta\, {\bf C}
\cdot \by-
\cos\beta\, {\bf H}_0\cdot\by). 
\end{equation}

Using (\ref{dyona}) and (\ref{dyonb}), we find 
completely regular
expressions for the functions ${\tilde\phi}(\br';\beta)$, 
${\tilde A}^a_i(\br';\beta)=\cos\beta {\da}^a_i
(\by=\br'\cos\beta)$, 
 ${\tilde A}^a_0(\br';\beta)=\mp\sin\beta {\bar\phi}
(\br';\beta)
+\alpha^a_0(\br';\beta)$ 
(see (\ref{ansa1})) immediately. If those 
are inserted into
(\ref{ansatz}), we have the explicit perturbative 
solution appropriate to a BPS dyon in the presence of 
uniform 
electromagnetic and Higgs field strengths asymptotically. 
Note that only elementary functions enter our solution 
(but in a rather
complicated way), and the result for ${\bf H}_0=0$ of
 course 
coincides with that already given in Ref. \cite{bak2}. 
Long-range
electromagnetic and Higgs fields, which are easily 
extracted from this
time-dependent solution to the field equations, again 
take simple forms. As for the ${\bf B}^{\rm em}$,  
${\bf E}^{\rm em}$
and $H^0$, the expressions given in   
(\ref{radfa}), (\ref{radfc}) and (\ref{radfd}) are still 
valid under the condition
that the acceleration parameter $\ba$ is now 
specified by (\ref{motioneq}).
On the other hand, the expression of ${\bf H}$ now 
contains also a 
uniform-field term over the result (\ref{radfc}), 
{\it viz.},
\begin{equation}
\label{radfcc}
{\bf H}(\br,t) \sim {\bf H}_0 +
{g_s\over 4\pi}{\hat{\bf R}\!-\!{\bf v}_{\rm ret}\over 
(1-\hat{{\bf R}}\cdot{\bf v}_{\rm ret})^3 R^2}
+\left\{ {g_s\over 4\pi}{(\hat{\bf R}\cdot \ba)\hat{\bf R}\over 
R}\right\}.
\end{equation}
This in turn implies that one may continue to
use the formula (\ref{radflux}), with $\ba$ 
specified by (\ref{motioneq}),
to find the radiated energy flux in the form of 
electromagnetic and Higgs waves.

\subsection{Electromagnetic and Higgs waves incident on dyons}

In Section III.1, the light scattering off a neutral 
BPS monopole was described in the long-wavelength 
limit. Since the 
theory  admits also a massless Higgs, one might also 
consider
a Higgs wave scattering by a BPS monopole/dyon, which 
would reveal tree-level
interactions between a massless Higgs and a BPS dyon. 
Therefore, 
to make our analysis complete, we 
will here analyze light and Higgs wave scattering by a 
BPS dyon with the help of
analogous perturbative scheme. 

In the presence of incident electromagnetic and Higgs 
plane waves, the dyon is 
expected to undergo a motion of
the form (\ref{motion}) with the vector $\ba$ describing 
the oscillating 
direction and amplitude of the dyon in responce to the 
incident waves.
The vector $\ba$ is taken to be real; this amounts to 
choosing the 
initial condition, ${\bf X}(0)=0$. Here ${\bf X}(t)$ 
describes the 
position ({\it i.e.} the center) 
of the dyon 
that is defined as the zero of the Higgs field 
$\phi(\br,t)$. We shall again
construct a solution to the field equations 
(\ref{fieldeq})-(\ref{pfieldeq}) 
corresponding to this 
oscillating dyon with incident electromagnetic and 
Higgs plane waves. Due to the 
oscillatory motion, it must radiate electromagnetic and 
Higgs waves as in the case
of a neutral monopole. Hence, the solution describes the 
scattering of 
light and Higgs particle by a
dyon. 

%We begin our analysis by writing down ansatz for the 
%solution:
One may begin the analysis with an ansatz for the solution:
\begin{eqnarray} 
\label{ascatan}
\phi^a(\br, t) &=&{\bar \phi}^a(\br-{\bf X};\beta) + 
{\rm Re}[{\tilde\Pi}^a (\br-{\bf X};\beta)e^{-i\omega t}]
%\nonumber
\\
%&=&{\bar \phi}^a(\br;\beta) + 
%{\rm Re}[\Pi^a (\br-{\bf X})e^{-i\omega t}]+O(a^2),\\
\label{bscatan}
A_\mu^a(\br, t) &=&
{\bar A}^a_\mu(\br-{\bf X};\beta) + 
{\rm Re}[\tilde\alpha^a_\mu (\br-{\bf X};\beta)
e^{-i\omega t}]
%\nonumber
%&=&{\bar A}^a_\mu(\br) + 
%{\rm Re}[\alpha^a_\mu (\br-{\bf X})e^{-i\omega t}]
+O(a^2),
\end{eqnarray} 
where (${\bar \phi}^a(\br;\beta)$, ${\bar A}^a_\mu
(\br;\beta)$) 
being the static
dyon solution characterized by  magnetic and 
electric charges
($g=\mp {4\pi/ e}$, $q=\mp {4\pi\tan\beta/ e}$). 
The functions 
$({\tilde\Pi}^a (\br-{\bf X};\beta), \tilde\alpha^a_\mu 
(\br-{\bf X};\beta))$
represent excitations from the undeformed,
but moving 
dyon with the center at 
${\bf X}(t)$, and especially contain  the asymptotic 
fields required for the
motion and the radiations emitted by the dyon. 
Despite of the clarity
in their interpretation, we shall not work with 
these functions due to
the complexity in resulting equations. Instead,
we define new functions 
\begin{eqnarray} 
\label{anewfunc}
{\tilde{\Pi'}}^a=
\tilde\Pi^a(\br-\bX;\beta)-
\tilde\bX\cdot\nabla \bar\phi^a(\br-\bX;\beta),\\
\label{bnewfunc}
{\tilde{\alpha'}}^a_\mu=
\tilde\alpha^a_\mu(\br-\bX;\beta)-
\tilde\bX\cdot\nabla \bar{A}^a_\mu(\br-\bX;\beta).
\end{eqnarray}
where $\tilde\bX$ is implicitly defined by the 
relation $\bX(t)={\rm Re}
[\tilde\bX e^{-i\omega t}]$.
These functions in fact 
represent the entire time-dependent corrections 
to the static
configurations.  
As in the case of a monopole, the functions 
($\tilde\Pi^a$, $\tilde\alpha^a_\mu$) and ($\tPi^a$, 
$\talpha^a_\mu$) are 
assumed to be $O(a)$ and we will solve the field
equations to the first order in $a$.
The field equation (\ref{fieldeq}) now reads
\begin{eqnarray} 
\label{anewfeq}
\!\!\!\!\!\!\!\!&&\!\!\!\!\!\!\!\!
(D_iD_i A_0)^a -i\omega (\bar{D}_i \talpha_0)^a= 
ie\omega\epsilon^{abc}\bar\phi^b\tPi^c -e^2 
\epsilon^{abc}\epsilon^{cde} \phi^b A_0^d \phi^e,\\
\label{bnewfeq}
\!\!\!\!\!\!\!\!&&\!\!\!\!\!\!\!\!(D_j G^{ij}\!)^a \!-\!
\omega^2 \talpha_i^a\! +\! i\omega (\bar{D}_i\talpha_0)^a
\!+\!2ie\omega\epsilon^{abc} \talpha^b_i \bar{A}_0^c\!=
\!e \epsilon^{abc}A_0^b (D_i A_0)^c 
\!-\!e \epsilon^{abc}\phi^b (D_i\phi)^c, 
\end{eqnarray} 
while the other field equation (\ref{pfieldeq}) becomes
\begin{eqnarray} 
\label{cnewfeq}
(D_i D_i\phi)^a \!+\!
\omega^2 \tPi^a
\!+\!2ie\omega\epsilon^{abc}  \bar{A}_0^b \tPi^c \!+\! 
ie\omega \epsilon^{abc}\talpha_0^b\bar\phi^c=e^2 
\epsilon^{abc}\epsilon^{cde} A_0^b A_0^d \phi^e,
\end{eqnarray} 
where only part of the relevant quantities are expressed 
in terms of ($\tPi^a$, $\talpha^a_\mu$). 

To proceed further, we  find it convenient to
introduce the functions $\tb^a_i(\br-{\bf X})$  by
\begin{equation}
\label{defbb}
 B^a_i(\br, t)= \mp{(D_i\phi)^a(\br,t)\over \cos\beta} 
-\tan\beta E^a_i
%[(D_k A_0)^a +i\omega \alpha^a_k]
\pm {\tb_i^a(\br-{\bf X})e^{-i\omega t}\over \cos\beta}.
\end{equation} 
Note that  $\tb^a_i$ effectively describe dynamical 
excitations from 
BPS saturated state satisfying combined Bogomol'nyi equation
(see (\ref{bogomol})),
\begin{equation}
\label{cbogo}
 B_k(\br, t)= \mp{(D_k\phi)^a(\br,t)\over \cos\beta}-
tan\beta E^a_k,
\end{equation} 

%Notice that this definition resembles a 
%combined Bogomol'nyi
%equation (see (\ref{bogomol})),
%\begin{equation}
%\label{cbogo}
% B_k(\br, t)= \mp{(D_k\phi)^a(\br,t)\over 
%\cos\beta}-tan\beta E^a_k,
%\end{equation} 
%and $b^a_i$ describes a correction to the 
%Bogomol'nyi equations. Hence
%a nonzero $b^a_i$ tell us that there are dynamic 
%excitations from the BPS
%saturated dyon state. 

If we use the relation  (\ref{defbb}) to eliminate 
$D_i\phi$ from 
(\ref{cnewfeq}) and the Bianchi identity $(D_iB_i)^a=0$, 
we obtain
\begin{eqnarray} 
\label{bdypert}
\omega^2\tPi_a=({\bar D}_i \tb_i)_a -ie\omega 
\epsilon_{abc}(\talpha^b_0{\bar\phi}^c +{\bar A}_0^b\tPi^c)
\end{eqnarray}
%(\ref{fieldeq})-(\ref{pfieldeq})
%are reduced to
while direct insertion of (\ref{defbb}) into 
(\ref{bnewfeq}) yields
\begin{eqnarray} 
\label{adypert}
\omega^2\talpha^a_i+i\omega ({\bar D}_i 
\talpha_0)^a &=&\mp
{1\over \cos\beta}\epsilon_{ijk}({\bar D}_j b_k)^a
+e 
\epsilon_{abc}\tb^b_i{\bar\phi}^c
\nonumber\\
&-&i\omega e \tan\beta 
\epsilon_{ijk}({\bar D}_j \talpha_k)^a -i\omega e
\epsilon_{abc}\talpha^b_i{\bar A}_0^c.
\end{eqnarray}  
One may also re-express the relation (\ref{defbb}) 
in terms 
of $\tPi^a$ and $\talpha^a_\mu$  as
\begin{equation}
\label{defbc}
 \epsilon_{ijk}({\bar D}_j \talpha_k)^a = 
\mp{({\bar D}_k\tPi)^a\over \cos\beta} 
\mp \cos\beta \epsilon_{abc}\talpha^b_i {\bar\phi}^c+
\tan\beta [
({\bar D}_i \talpha_0)^a -i\omega \talpha^a_i]\pm
{\tb_i^a\over \cos\beta}.
\end{equation} 
It is then not difficult to verify 
that Eq.~(\ref{anewfeq}) is identically satisfied 
when (\ref{defbc}), (\ref{bdypert}) and (\ref{adypert}) 
are used.  

Taking an appropriate combination of (\ref{defbc}) and 
(\ref{adypert}) to
eliminate the $\talpha^a_i$-dependence  
and using the relation (\ref{bdypert}), 
we can derive a second-order equation for $\tb^a_i$,  
which reads
\begin{equation}
\label{dyhelmholtz}
[({\bar D}_k{\bar D}_k+\omega^2) \tb_i]_a+e^2\cos^2\beta 
\epsilon_{abc}\epsilon_{bde} \tb_i^d{\bar\phi}^e
{\bar\phi}^c=0.
\end{equation}
On the other hand, eliminating the  
$\epsilon_{ijk}({\bar D}_j \talpha_k)^a$ terms
from  (\ref{defbc}) and  (\ref{adypert}) leads to
\begin{eqnarray} 
\label{cdypert}
\omega^2\talpha^a_i =
\mp\cos\beta\epsilon_{ijk}({\bar D}_j \tb_k)^a
+e 
\cos^2\beta\epsilon_{abc}\tb^b_i{\bar\phi}^c 
-i\omega ({\bar D}_i \talpha_0)^a  
\pm i\omega \sin\beta[({\bar D}_i \tPi)^a +\tb^a_i].
\end{eqnarray}
Once $\tb^a_i$ are obtained from (\ref{dyhelmholtz}), 
we may use  
(\ref{bdypert}) and  (\ref{cdypert}) to fix 
($\tPi^a$, $\talpha^a_i$) 
up to  unknown functions $\talpha^a_0$. Again note that
there is no equation for $\talpha^a_0$, which merely 
reflects that 
the choice of $\talpha^a_0$ is related to  pure gauge 
degrees of freedom. The equation (\ref{dyhelmholtz})
is the same as (\ref{helmholtz}) when we scale $\br$ 
to ${\br/\cos\beta}$,
and $\omega$ to $\omega\cos\beta$. Thus the scattering 
solution 
immediately
follows  if we use the results of Section III.1:
\begin{equation}
\label{bsolution}
 \tb_{i}^a =\pm i\omega^2 a_i f \coth m_v r'\cos\beta 
e^{i {\bf k}\cdot \br}{{\hat {r'}}}^a \mp i\omega^2 a_i
{e^{i\omega r}\over er'\cos\beta}{\hat {r'}}^a
+ O(a\omega^3).
\end{equation}
where $\br'\equiv\br-{\bf X}$. (We will see 
below that this particular 
homogeneous solution in fact describes 
the oscillating dyon by incident
electromagnetic and Higgs plane-waves. Of course, 
the solution is not the most
general solutions of the above equation.) Upon 
making the gauge  choice 
\begin{equation}
\label{gauge}
\talpha^a_0 =\mp \sin\beta \tPi^a + \omega a_i 
{\bar A}^a_i,
\end{equation}
and using (\ref{bdypert}) and  (\ref{bdypert}), 
we find the expressions
\begin{eqnarray} 
\label{pialasym}
&&\tPi^a \sim \mp \omega 
[{\bf a}\cdot{\bf \hat k}f e^{i\bk\cdot\br}-
{\ba\cdot {\hat\br}\over er\cos\beta}e^{i\omega r}]
{\hat r}^a,\\
&&\talpha^a_i\sim  \omega f 
[({\bf \hat k}\times {\bf a})_i \cos\beta  -
a_i \sin\beta]  e^{i\bk\cdot\br}{\hat r}^a
-\omega [({\bf \hat r}\times {\bf a})_i \cos\beta  
-a_i \sin\beta]  
e^{i\omega r}{\hat r}^a
\end{eqnarray}
in the scattering region where the terms of 
$O(r^{-2})$ are ignored.
Consequently, the electromagnetic and Higgs fields 
in the asymptotic region are 
given as
\begin{eqnarray} 
\label{BEasymp}
{\bf B}^{\rm em}&=& \mp i\omega^2 
[\cos\beta {\bf \hat k}\times ({\bf \hat k}\times 
{\bf a})-\sin\beta
({\bf \hat k}\times {\bf a})]  f e^{i\bk\cdot\br 
-iwt}\nonumber\\
&\mp& i\omega^2 
[\cos\beta {\bf \hat r}\times ({\bf \hat r}\times 
{\bf a})-\sin\beta
({\bf \hat r}\times {\bf a})]  { e^{i\bk\cdot\br -
iwt}\over er\cos\beta}\\
{\bf E}^{\rm em}&=& \mp i\omega^2 
[\cos\beta ({\bf \hat k}\times {\bf a}) +\sin\beta
{\bf \hat k}\times ({\bf \hat k}\times {\bf a})
]  f e^{i\bk\cdot\br -iwt}\nonumber\\
&\mp& i\omega^2 
[\cos\beta ({\bf \hat r}\times {\bf a})+\sin\beta 
{\bf \hat r}\times 
({\bf \hat r}\times {\bf a})
]  { e^{i\bk\cdot\br -iwt}\over er\cos\beta}\\
\label{HHasymp}
{\bf H}\ \ &=& -i\omega^2 
({\bf a}\cdot{\hat {\bf k}}){\hat{\bf k}}f 
e^{i\bk\cdot\br-i\omega t}+i\omega^2
(\ba\cdot {\hat\br}) {\hat \br}{e^{i\omega r-i\omega t}
\over  er\cos\beta},\\
{\bf H}_0\ &=& -i\omega^2 
({\bf a}\cdot{ \hat {\bf k}})f e^{i\bk\cdot\br-
i\omega t}+i\omega^2
(\ba\cdot {\hat\br}){e^{i\omega r-i\omega t}\over  
er\cos\beta},
\label{H0asymp}
\end{eqnarray}  
where only the real parts are relevant. {}From  
those expressions,
% for the field strengths, 
one may clearly see the presence of incident 
plane-waves as well
as the electromagnetic and Higgs radiation fields 
emitted by the dyon. As expected, 
the force law can be verified explicitly by finding 
zero of $\phi(\br,t)$: 
\begin{equation}
\label{scatforce}
M {\ddot {\bf X}}=M{d^2\over dt^2}{\rm Re}[-i\ba 
e^{-i\omega t}]
={\rm Re}[g{\bf B}^{\rm em}_{\rm inc}+
q{\bf E}^{\rm em}_{\rm inc}+g_s{\bf H}^{\rm em}_{\rm inc}
]_{\br={\bf X}}.
\end{equation} 
Here the subscript {\it inc} indicates that it 
refers only to
the incident part of the given field. 
The results (\ref{BEasymp})-(\ref{H0asymp}) can be 
used to calculate the related
scattering cross sections.
With the energy momentum tensor  (\ref{tensor}), 
the time-averaged
incident flux densities in  electromagnetic and 
Higgs sectors are 
respectively
\begin{eqnarray}
\label{incflux}
(T^{0i})^{\rm em}_{\rm inc}&=& 
{1\over 2} \omega^4 f^2 |\ba\times\hat{\bf k}|^2 {\hat k}_i,\\
(T^{0i})^{\rm Higgs}_{\rm inc}&=& 
{1\over 2} \omega^4 f^2 |\ba\cdot\hat{\bf k}|^2 {\hat k}_i,
\end{eqnarray}
while the time-averaged radiated energy flux densities are
\begin{eqnarray}
\label{raddyflux}
(T^{0i})^{\rm em}_{\rm rad}&=& 
{ \omega^4 \over 2 e^2 r^2\cos^2\beta} 
|\ba\times\hat{\bf r}|^2 {\hat r}_i,\\
(T^{0i})^{\rm Higgs}_{\rm rad}&=& 
{ \omega^4 \over 2 e^2 r^2\cos^2\beta}
|\ba\cdot\hat{\bf r}|^2 {\hat r}_i.
\label{aaraddy}
\end{eqnarray}
Based on these, we find that, when a light is 
incident upon the dyon, {\it i.e.},
$\ba\cdot{\bf k}=0$, the related differential 
cross sections are\footnote{In view of the relation 
$g^2+q^2=g_s^2$, the multiplicative
factors appearing on the right hand sides of 
(\ref{dyemcross}) and (\ref{adyemcross})
are actually the same; here [and also in 
(\ref{dyhiggscross}) 
and (\ref{adyhiggscross})], 
we have just written the expression in such a 
way that the vertices involved
may clearly be seen.}
%identified as
%(note that $g^2+q^2=g_s^2$)
\begin{eqnarray} 
\label{dyemcross}
&&\left({d\sigma\over d\Omega}\right)_{{\rm em}
\rightarrow {\rm em}}=
\left({g^2+q^2\over 4\pi M}\right)^2 \sin^2\Theta,\\
\label{adyemcross}
&& \left({d\sigma\over d\Omega}\right)_{{\rm em}
\rightarrow {\rm Higgs}}=
\left({g^2+q^2\over 4\pi M}\right)\left({g_s^2\over 
4\pi M}\right) \cos^2\Theta
\end{eqnarray}
where $\Theta$ is the angle between $\hat\br$ and 
the combination 
$g{\bf B}^{\rm em}_{\rm inc}+
q{\bf E}^{\rm em}_{\rm inc}$. On the other hand, 
for an incident 
Higgs wave, we find 
\begin{eqnarray} 
\label{dyhiggscross}
&&\left({d\sigma\over d\Omega}\right)_{{\rm Higgs}
\rightarrow {\rm em}}=
\left({g_s^2\over 4\pi M}\right)\left({g^2+q^2\over 
4\pi M}\right) 
\sin^2\theta, \\
\label{adyhiggscross}
&& \left({d\sigma\over d\Omega}\right)_{{\rm Higgs}
\rightarrow {\rm Higgs}}=
\left({g_s^2\over 4\pi M}\right)^2 \cos^2\theta
\end{eqnarray} 
where $\theta$ is the angle between $\hat\br$ and 
$\hat{\bf k}$.

As should be the case,
% with vanishing $\beta$, 
the cross sections 
(\ref{dyemcross}) and (\ref{adyemcross}) for 
vanishing $\beta$ agree with those of light 
scattering by a monopole in 
(\ref{crossection}) and (\ref{acrossection}). [But 
the case of  
the Higgs and dyon/monopole
scattering was not considered before.]
%collision,  whose crossections are  also given in 
%(\ref{dyhiggscross})-(\ref{adyhiggscross}) 
%with $\beta=0$, 
%was not previously
%analyzed in Ref.~\cite{bak1}. 
Also it should be stressed that the cross 
sections found above
are manifestly duality-symmetric ({\it i.e.}, 
involve the combination 
$g^2+q^2$ only)
and have the same form as the corresponding cross 
sections for a W-particle
(see (\ref{wcrossection})). In fact the formulas 
(\ref{dyemcross})-(\ref{adyhiggscross})
apply to solitons and elementary quanta alike, only if
appropriate values for mass and various charges are used.

\section{Effective Theory for Electromagnetic 
and Higgs Scalar
Interactions of BPS Dyons}
\subsection{Duality-invariant Maxwell theory}

According to the results of the preceding section, 
behaviors 
of BPS dyons in low-energy processes are not very different 
from those of W-particles; that is, solitons and elementary
field quanta behave alike. This in turn suggests that there 
should exist a simple effective field theory for low-energy
BPS dyons interacting with long-range fields. But, unlike
W-particles,
dyons carry both electric and magnetic charges and so their 
electromagnetic interactions cannot be accounted for by the 
usual Maxwell theory---we need a duality-symmetric 
generalization
of the latter. Even from sixties, Schwinger considered such
duality-symmetric Maxwell theory seriously\cite{schwinger}, 
and then several 
different versions were developed by him and 
others\cite{zwanziger} 
since.
For our discussion we find the simple first-order 
action approach,
given by Schwinger\cite{schwinger2} in 1975, 
adequate. Its basic idea will
be recalled briefly in this subsection.

The goal is to find a simple Lagrangian description
for the generalized Maxwell system
\begin{eqnarray} 
\label{maxwell}
\partial_\nu F^{\nu\mu}= 
J_e^{\mu}(x),\ \ \ \ \  \partial_\nu\!\,^*\!F^{\nu\mu}= 
J_g^{\mu}(x) 
\end{eqnarray} 
where $^*\!F^{\mu\nu}={1\over 2}
\epsilon^{\mu\nu\lambda\delta}F_{\lambda\delta}$, 
and $J_e$ and $J_g$ denote conserved electric and 
magnetic sources, respectively.
This system is marked by the duality symmetry
\begin{eqnarray} 
&&{J'}_e^{\mu}(x)= \cos \alpha\, J^\mu_e (x)+\sin 
\alpha\, J^\mu_g (x),
\ \ \ {J'}_g^{\mu}(x)= -\sin\alpha\, J^\mu_e (x)+
\cos \alpha\, J^\mu_g (x),
\nonumber\\
&& {F'}^{\mu\nu}(x)=\cos\alpha\, F^{\mu\nu}(x)+
\sin\alpha\,^*\! F^{\mu\nu}(x)
\label{duality}
\end{eqnarray}
For a given distribution of $J^\mu_e$ and $J^\mu_g$, 
the
field strengths $F^{\mu\nu}$ 
(satisfying suitable asymptotic conditions) can 
be determined using
(\ref{maxwell}). But, for a Lagrangian, vector 
potentials are needed.
Based on the second of (\ref{maxwell}), we may 
here introduce 
the vector potential $A^\mu(x)$ by
\begin{eqnarray} 
\label{amu}
F^{\mu\nu}(x)=\partial^\mu A^\nu(x)\!-\!\partial^\nu 
A^\mu(x)
\!-\!\int d^4 x' (n\cdot \partial)^{-1}(x,x')
{1\over 2}\epsilon^{\mu\nu\lambda\delta}[n_\lambda 
J_{g\delta}(x')
\!-\!n_\delta J_{g\lambda}(x')].
\end{eqnarray} 
Here, $n^\mu$ may be any fixed, spacelike, unit vector, 
and Green's 
function $(n\cdot \partial)^{-1}$ is realized by
\begin{eqnarray} 
\label{kernel}
(n\cdot \partial)^{-1}(x,x')&=&\int^\infty_0 d\xi[a 
\delta^4(x-x'-n\xi)
-(1-a) \delta^4(x-x'+n\xi)]\nonumber\\
&=&\left\{ a\Theta [n\cdot (x-x')]- (1-a)\Theta 
[-n\cdot (x-x')]
\right\}\delta_n(x-x'),
\end{eqnarray} 
where one can choose either $a=0,1$ (semi-infinite string) 
or 
$a=1/2$ (symmetric infinite string), and 
$\delta_n(x-x')$ denotes a 
3-dimensional $\delta$-function with a support 
on the hypersurface
orthogonal to $n^\mu$.
Similarly, the first of (\ref{maxwell}) informs us 
that we may also write
\begin{eqnarray} 
\label{cmu}
F^{\mu\nu}(x)=-{1\over 2}\epsilon^{\mu\nu\lambda\delta}
(\partial_\lambda 
C_\delta(x)\!-\!\partial_\delta C_\lambda(x))
\!+\!\int d^4 x' (n\cdot \partial)^{-1}(x,x')
[n^\mu J_{e}^\nu(x')
\!-\!n^\nu J_{e}^\mu(x')],
\end{eqnarray} 
$C^\mu(x)$ being another vector potential which is 
unrestricted 
by the first equation of (\ref{maxwell}) alone.

The two potentials $A^\mu$ and $C^\mu$ cannot be 
completely
independent, being connected through (\ref{amu}) 
and (\ref{cmu}).
In fact the latter relations allow one to determine 
the potentials 
in terms of $F^{\mu\nu}$, up to a gauge transformations 
separately for  $A^\mu$ and $C^\mu$. Explicitly, 
we have 
\begin{eqnarray} 
\label{amusol}
A^{\mu}(x)&=&-\int d^4 x' (n\cdot \partial)^{-1}(x,x')
n_\nu\, F^{\mu\nu}(x')+\partial^\mu \Lambda_e(x),
\\
\label{cmusol}
C^{\mu}(x)&=&-\int d^4 x' (n\cdot \partial)^{-1}(x,x')
n_\nu\,^*\! F^{\mu\nu}(x')+\partial^\mu \Lambda_g(x),
\end{eqnarray} 
where $\Lambda_e(x)$ and $\Lambda_g(x)$ are 
arbitrary
gauge functions (which may be set to zero in the 
gauge $n_\mu A^\mu(x)=
n_\mu C^\mu(x)=0$). Because of these, we can 
regard the potential 
$C^\mu$ to represent the field-strength-dependent 
function $C_\mu(F)$
as specified by (\ref{cmusol}) while the field 
strengths $F^{\mu\nu}$
are expressed in terms of the potential $A^\mu$ 
through (\ref{amu})\footnote{Alternatively, utilizing the 
relations 
(\ref{amusol}) and (\ref{cmu}), one may assign 
a primary role on the 
{\it dual} potential $C_\mu$ (rather than $A_\mu$).}.
We also remark that, with the choice $n^\mu=(0, 
\hat{\bf n})$ 
%and $a=1$ 
(see (\ref{kernel})), using the formula 
(\ref{amusol}) (for $\Lambda_e(x)=0$) 
with the magnetic
Coulomb field of a point monopole leads 
to the famous Dirac vector
potential with a semi-infinite string 
along the direction 
$\hat{\bf n}$  if the value $a=0,1$ is 
assumed in the Green's function
realization (\ref{kernel}). Varying the 
direction of $\hat{\bf n}$
just leads to gauge equivalent potentials 
if the magnetic charge 
carried by the monopole satisfies the 
well-known quantization 
condition\cite{dirac}. 
%or $-\hat{\bf n}$. 
On the other hand, if one adopts the Schwinger 
value\cite{schwinger,schwinger2}
$a=1/2$ in (\ref{kernel}), the resulting 
monopole vector potential
will contain a symmetrically-located infinite 
string singularity
along the direction $\pm \hat{\bf n}$. 
In the latter case,
the vector potentials corresponding to 
different choices 
of $\hat{\bf n}$ can be shown to be gauge 
equivalent\cite{schwinger,schwinger2}
if the magnetic charge is quantized by twice the Dirac unit.
As for the magnetic monopoles of the 
Yang-Mills-Higgs system,
either value of $a$ may be adopted to define 
$(n\cdot \partial)^{-1}$;
but, if one wishes to have a manifestly duality-symmetric 
action formulation, the Schwinger value $a=1/2$ may 
be chosen (see below).

We are now ready to present Schwinger's first-order 
action
approach. It is based on the action
\begin{eqnarray} 
\label{schwinger}
S=\int d^4 x\left\{ {1\over 4} F^{\mu\nu}F_{\mu\nu}
-{1\over 2}F^{\mu\nu}(\partial_\mu A_\nu-\partial_\nu A_\mu)
-J_e^\mu A_\mu -J^\mu_g C_\mu(F)\right\},
\end{eqnarray}  
where $A_\mu$
and $F^{\mu\nu}$ are taken to be independent 
fields and $C_\mu(F)$
are specified as above, {\it i.e.}, through (\ref{cmusol}).
Obviously, the first Maxwell equation $\partial_\nu F^{\nu\mu}= 
J_e^{\mu}(x)$
is the consequence of $\delta S/\delta A_\mu(x)=0$.
On the other hand, from $\delta S/\delta F_{\mu\nu}(x)=0$,
we obtain
\begin{eqnarray} 
\label{amuaa}
F^{\mu\nu}(x)=\partial^\mu A^\nu(x)\!-\!\partial^\nu A^\mu(x)
\!+\!\int d^4 x'
{1\over 2}\epsilon^{\mu\nu\lambda\delta}[n_\lambda 
J_{g\delta}(x')
\!-\!n_\delta J_{g\lambda}(x')]  (n\cdot \partial)^{-1}(x',x),
\end{eqnarray} 
or taking the dual,
\begin{eqnarray} 
\label{amudual}
^*\!F^{\mu\nu}(x)={1\over 2}\epsilon^{\mu\nu\lambda\delta}
(\partial_\lambda A_\delta(x)\!-\!\partial_\delta A_\lambda(x))
\!-\!\int d^4 x'
[n^\mu J_{g}^\nu(x')
\!-\!n^\nu J_{g}^\mu(x')]  (n\cdot \partial)^{-1}(x',x).
\end{eqnarray} 
Then, based on (\ref{amudual}), it is easy to derive
the second Maxwell equation $\partial_\nu\,^*\!F^{\nu\mu}= 
J_g^{\mu}(x)$ also. Therefore, the action (\ref{schwinger})
can be used to describe the system (\ref{maxwell}). 
Here notice 
another consequence
of (\ref{amuaa}): multiplying (\ref{amuaa}) by $n_\nu$
and picking the gauge $n_\mu A^\mu=0$ yields
\begin{eqnarray} 
\label{qqq}
n_\nu F^{\mu\nu}(x)=- (n\cdot \partial) A^\mu(x),
\end{eqnarray} 
and hence the relation (\ref{amusol}) follows. 
Moreover, our definition of $C^\mu(F)$ and the 
first Maxwell
equation 
$\partial_\nu F^{\nu\mu}= 
J_e^{\mu}(x)$ may be used to confirm the 
representation (\ref{cmu}).

Astute readers should have noticed that (\ref{amuaa}) is 
not quite our earlier equation (\ref{amu}), unless 
our Green's
function $(n\cdot \partial)^{-1}(x',x)$ satisfies 
the symmetry
property
\begin{eqnarray} 
\label{symm}
 (n\cdot \partial)^{-1}(x',x)=-(n\cdot \partial)^{-1}(x,x')\,.
\end{eqnarray} 
Actually, this odd character of Green's function 
is also necessary
for the action (\ref{schwinger}) to be invariant
under the duality transformation (\ref{duality})
(now generalized to include the duality rotation between 
$A_\mu$ and $C_\mu(F)$ in an obvious way)\cite{schwinger2}.
The condition (\ref{symm}) is met if the Schwinger
value $a={1\over 2}$ is chosen
with our representation (\ref{kernel}).
% Note that, 
%with the choice
%$a={1\over 2}$ made, the (static)
%monopole vector potential given by (\ref{amusol}) 
%(with $\Lambda_e(x)=0$)
%has a symmetrically located infinite string. 

\subsection{Low-energy effective theory of BPS dyon}

Our detailed analysis of nonlinear field equations
(given in Sec. III) revealed that BPS dyons behave just 
like point-like objects carrying electric, magnetic 
and dilaton charges.
[This does not mean that core region 
%({\it i.e.}, $r 
%{\stackrel{\textstyle <}{\sim} 1/m_v$) 
of the dyon profile remains rigid; rather, the 
core profile gets 
deformed suitably to accommodate any change 
in the 
long-range tail part.] 
This observation applies to our force law
 (\ref{dycolo}), to the 
near-zone and radiation-zone fields given in 
(\ref{radfa})-(\ref{radfd})
and (\ref{radfcc}), and to the scattered waves 
of electromagnetic and 
Higgs particles found in (\ref{BEasymp})-(\ref{H0asymp}). As 
a matter of fact, these results are exact 
parallels of the 
corresponding formulas
for the W-particles,
aside from  the ubiquitous sign of duality-invariant
electromagnetic coupling  in all of our formulas 
derived for BPS dyons.  Therefore, we should be able to
account for the entire low-energy dynamics 
involving $N$ BPS dyons
and massless fields by a simple effective field theory,
described by an action corresponding to a duality-symmetric 
generalization of the low-energy W-particle  
action (\ref{lageff}).
We shall make this statement more precise below.

What we ask for our effective field theory is that it should be able
to describe to a good approximatiom the dynamical developement of a
configuration of $N$ well-separated BPS dyons ({\it i.e.}, at any
given instant, the Higgs field has $N$ zeros at various locations),
while allowing incoming and outgoing radiations (with moderate
frequency) of massless fields.  For this purpose, we must first
specify appropriate dynamical variables which may enter our effective
theory. We shall here keep the position coordinates of BPS dyons (or
the location of zeros in the Higgs field), ${\bf X}_n (t)\,
(n=1,\cdot\cdot\cdot, N)$, the electromagnetic fields ($A_\mu(x)$),
and the Higgs field ($\varphi(x)$). Each BPS dyon has three kinds of
charges, that is, $q_n$, $g_n(=\mp 4\pi/e)$ and
$(g_s)_n(=\sqrt{g_n^2+q_n^2})$ for the n-th dyon; these charges are
made local sources for the electromgnetic and Higgs fields. The
electromagnetic field strength $F^{\mu\nu}$ may be defined so that
(\ref{amu}) may hold, and the Higgs field strength by
$H_\mu=-\partial_\mu\varphi$.  In our perturbative solutions given in
Sec.III, how should one identify the contributions which may duly be
associated with the fields $A^\mu$ and $\varphi$ (or the field
strengths $F^{\mu\nu}$ or $H_\mu$)?  Actually, for all of our explicit
solutions, the field strengths $G^{\mu\nu}_a$ and $(D_\mu\phi)^a$ in
the region away from the dyon core ({\it i.e.}, for $m_v r \gg 1$)
have nonvanishing components only in the {\it isospin} direction
$\hat{\phi}^a$. Only the fields in this region are relevant for the
present discussion and here one may identify $A^\mu$ and $\varphi$
unambiguously by going to the unitary 
gauge\footnote{Gauge-invariant identification can also be given.
Clearly, in the region away from the core, we may set $\varphi \simeq
|\phi|-f$, which in turn leads to $H_\mu\simeq
-\hat{\phi}^a (D_\mu\phi)^a$. Also note that $\hat{\phi}^a
G^{\mu\nu}_a$ in this region is essentially the same as the
gauge-invariant 't Hooft tensor\cite{thooft}, $F^{\mu\nu}=\hat{\phi}^a
G^{\mu\nu}_a -{1\over e}\epsilon^{abc} \hat{\phi}^a(D^\mu\hat{\phi})^b
(D^\nu\hat{\phi})^c$, which is known to satisfy the generalized
Maxwell equation (\ref{maxwell}).  Using this 't Hooft tensor, one may
then simply define the electromagnetic field $A^\mu$, say, by the
relation (\ref{amu}).}, that is, $\phi^a(x)=(f+\varphi(x))
\delta_{a3}$ and $A^\mu(x)=A^\mu_3(x)$ away from the core region.
Fields within the dyon core and charged vector fields correspond to
the freedoms to be integrated out.

We are now ready to write down the action,
which incorporates all of our findings on low-energy 
processes involving
BPS dyons. Noting that the results of our analysis for 
the dyons differ from
those for W-particles only by the presence of the 
electromagnetic
duality symmetry, the desired low-energy action is  
given by the form
%(\ref{lageff}), but with $L_{\rm eff}$ now changed as
\begin{eqnarray}
\label{e413}
S_{\rm eff}&=&\int d^4 x \biggl\{{1\over 4} F^{\mu\nu}
F_{\mu\nu}-{1\over 2}F^{\mu\nu}
(\partial_\mu A_\nu -\partial_\nu A_\mu)
-{1\over 2} \partial_\mu \varphi \partial^\mu 
\varphi\biggr\}\nonumber\\ 
&+&
\int dt \sum_{n=1}^N \biggl\{\,-\!\left(M_n + 
(g_s)_n \varphi(\bX_n,t)\right)
\sqrt{1-{\dot \bX}_n^2}\biggr.\nonumber\\
& -& \biggl.
\!q_n [A^0(\bX_n,t)\!-\!{\dot\bX}_n\cdot {\bf A}
(\bX_n,t)] -\!g_n [C^0(\bX_n,t)\!-\!{\dot\bX}_n\cdot 
{\bf C} (\bX_n,t)]\,\biggr\},
\end{eqnarray}  
where
$C^\mu=(C^0, {\bf C})$, as a function of $F^{\mu\nu}$,
are defined by (\ref{cmusol}) with Green's functions
$(n\cdot \partial)^{-1}$ satisfying the symmetry 
property (\ref{symm}).
As one can easily verify, the above action is
still invariant under the scale transformation of the 
form (\ref{gaugetr}).
Considering the variations of $F^{\mu\nu}$ and $A^\mu$, 
we then obtain 
(\ref{amu}) and the generalized Maxwell equations 
(\ref{maxwell})
with the source term given by
\begin{eqnarray} 
\label{e414}
 &&J^0_g(x)=\sum^N_{n=1} g_n \delta^3 ({\bf x}-\bX_n(t)),
\ \ \ 
{\bf J}_g(x)=\sum^N_{n=1} g_n \dot{\bX}_n (t)
\delta^3 ({\bf x}-\bX_n(t)),
\nonumber\\
 &&J^0_e(x)=\sum^N_{n=1} q_n \delta^3 ({\bf x}-\bX_n(t)),
\ \ \ 
{\bf J}_e(x)=\sum^N_{n=1} q_n \dot{\bX}_n (t)
\delta^3 ({\bf x}-\bX_n(t)).
\end{eqnarray} 
The corresponding equation of motion
for the field $\varphi$ reads
\begin{eqnarray} 
\label{e415}
\partial_\mu\partial^\mu \varphi (x)= \sum^N_{n=1}
(g_s)_n \sqrt{1-\dot{\bf X}_n } \delta^3 ({\bf x}-\bX_n (t))
\equiv J_s (x)\,.
\end{eqnarray} 
On the other hand, the $\bX_n$-variation with our 
action leads to 
the equation of motion
\begin{eqnarray}
\label{e416}
&&{d\over dt}\left[\left\{m_v\! +\!(g_s)_n 
\varphi(\bX_n,t)\right\}{{\bf V}_n\over 
\sqrt{1\!-\!{\bf V}_n^2}}\right]=
q_n [ {\cal F}^{0i}(\bX_n,t) +{V}_n^j 
{\cal F}^{ij}(\bX_n,t)]\nonumber\\
&&\ \ \ \ \ \ +g_n [ \bar{{\cal F}}^{0i}(\bX_n,t) +{V}_n^j 
\bar{{\cal F}}^{ij}(\bX_n,t)]
+(g_s)_n \sqrt{1\!-\!{\bf V}_n^2} \nabla\varphi(\bX_n, t),
\end{eqnarray} 
where we have defined ${\cal F}^{\mu\nu}\equiv 
\partial^\mu A^\nu-\partial^\nu
A^\mu$ and $\bar{{\cal F}}^{\mu\nu}\equiv 
\partial^\mu C^\nu-\partial^\nu
C^\mu$. Here, because of (\ref{amu}) 
and (\ref{cmu}), we have 
${\cal F}^{\mu\nu}= 
{ F}^{\mu\nu}$ and $\bar{\cal F}^{\mu\nu}=\,^*\! 
F^{\mu\nu}$ 
almost everywhere, that is, away from the string 
singularity;
in this way, the force law (\ref{dycolo}) is also 
incorporated in our action.
The effective theory defined by the above action, 
by its very construction, will reproduce all the 
consequences in Sec.III
in the proper kinematical regime.

When BPS dyons in the system are sufficiently 
slow-moving so that only 
negligible radiations are produced, the above 
effective field theory 
may be turned into the effective particle 
Lagrangian analogous to 
(\ref{modlag}).  For this, it suffices to integrate out
the fields $A^\mu(x)$ and $\varphi(x)$ using the 
near-zone solutions to
the respective equations of motion (for a given 
distribution of sources $J^\mu_g(x)$, 
$J^\mu_e(x)$ and $J_s(x)$; this is the same 
procedure to obtain the 
slow motion Lagrangian (\ref{modlag}) 
for W-particles (see also Appendix A). 
Then the Higgs field is expressed as (see (\ref{hpotential}))
\begin{eqnarray} 
\varphi (\bx,t)
&=& - {1\over 4\pi} \sum_n {(g_s)_n\sqrt{1-\dot\bX^2_n}
\over |\bx-\bX_n|}
+{1\over 4\pi}
{\partial\over\partial t} \left(\sum_n 
{(g_s)_n\sqrt{1-\dot\bX^2_n}}\right)\nonumber\\
&-&{1\over 8\pi} {\partial^2\over\partial t^2} 
\left( \sum_n (g_s)_n \sqrt{1-\dot\bX^2_n}\,
|\bx-\bX_n| \right)
+ \cdot\cdot\cdot\ \ .
\label{e417}
\end{eqnarray}
To obtain the corresponding expression for $A^\mu(x)$, 
one may use the formula
(\ref{kernel}) with help of the following expression for 
$F^{\mu\nu}$ (describing the 
near-zone solution to the generalized Maxwell equation 
(\ref{maxwell}):
\begin{eqnarray} 
\label{e418}
&&F^{0i} (\bx,t)
= {1\over 4\pi} \sum_n {q_n (x^i-X^i_n)\over 
|\bx-\bX_n|^{3/2}}
\left[ 1-
{3\over 2} {(\bx-\bX_n)\cdot \dot{\bX}_n\over |\bx-\bX_n|}+
{1\over 2}\dot\bX^2_n\right]\nonumber\\
&&\ \ \ \ \ \ \ \ -{1\over 4\pi}
\sum_n {g_n\epsilon^{ijk} \dot{X}^j_n(t)  (x^k-X^k_n)\over 
|\bx-\bX_n|^{3/2}} + O(\dot{X}^3) \\
\label{e419}
&&{1\over 2}\epsilon^{ijk}F_{jk}(\bx,t)\!=\!{1\over 4\pi} 
\sum_n\!\! {q_n\epsilon^{ijk} \dot{X}^j_n  
(x^k\!-\!X^k_n)\over 
|\bx-\bX_n|^{3/2}}
\!+\! {1\over 4\pi}\! \sum_n\!\! {g_n (x^i\!-\!X^i_n)\over 
|\bx\!-\!\bX_n|^{{3\over 2}}}
\!+\!O(\!\dot{X}^2\!). 
\end{eqnarray}
Given this expressions and the choice $n^\mu=(0, \hat{n})$, 
the integral in the right hand side of (\ref{kernel})
may be performed to discover, modulo gauge transformation, 
the following (near-zone) expression for 
the field $A^\mu(x)$:
\begin{eqnarray} 
A^0 (\bx,t)\!&=&\! 
\sum_n \left[  {q_n\over 
4\pi|\bx\!-\!\bX_n|}\!+\!{q_n\over 8\pi} 
{\partial^2\over\partial t^2}
 |\bx\!-\!\bX_n| \!+\!g_n \dot\bX_n\cdot 
{\bf \omega}(\bx;\bX_n)
\right]\!+\!O(\!\dot{X}^3\!)
\label{e420}
\\
A^i (\bx,t)\!&=&\! 
\sum_n \left[  {q_n\dot\bX^i_n\over 
4\pi|\bx-\bX_n(t)|}+g_n\omega^i(\bx;\bX_n)
\right]+O(\dot{X}^2)
\label{e421}
\end{eqnarray}
where $\omega^i(\bx;\bX_n)$ denotes the unit-monopole static 
vector potential (with a symmetrically-located 
infinite string), given by)
\begin{eqnarray} 
\omega (\bx;\bX_n)=-{1\over 8\pi}\left[  {\hat{n}\times 
(\bx-\bX_n)/|\bx-\bX_n|\over 
|\bx\!-\!\bX_n|-\hat{n}\cdot (\bx-\bX_n)}-
{\hat{n}\times 
(\bx-\bX_n)/|\bx-\bX_n|\over 
|\bx\!-\!\bX_n|+\hat{n}\cdot (\bx-\bX_n)}
\right]
\label{e422}
\end{eqnarray}
Note that the electric charge contributions in 
(\ref{e420}) and 
(\ref{e421}) are identical to those in 
(\ref{apotential}) and 
(\ref{bpotential}).  Also required is the expression
for the magnetic potential $C^\mu$. Using (\ref{e418}) and 
 (\ref{e419}) in (\ref{cmusol}) and making 
appropriate gauge 
transformation, one has an expression dual to (\ref{e420})
and (\ref{e421}):
\begin{eqnarray} 
C^0 (\bx,t)\!&=&\! 
\sum_n \left[  {g_n\over 
4\pi|\bx\!-\!\bX_n|}\!+\!{g_n\over 8\pi} 
{\partial^2\over\partial t^2}
 |\bx\!-\!\bX_n| \!-\!q_n \dot\bX_n\cdot 
{\bf \omega}(\bx;\bX_n)
\right]\!+\!O(\!\dot{X}^3\!)
\label{e423}
\\
C^i (\bx,t)\!&=&\! 
\sum_n \left[  {g_n\dot\bX^i_n\over 
4\pi|\bx-\bX_n|}-q_n\omega^i(\bx;\bX_n)
\right]+O(\dot{X}^2)
\label{e424}
\end{eqnarray}

The desired effective Lagrangian will result
if the fields $A^\mu(x)$ and $\varphi(x)$ are eliminated from
the action (\ref{e413}) by using the above effective solutions.
Here it is useful to notice that, thanks to the 
field equations satisfied by  $A^\mu$ and $\varphi$,
contributions from the massless field action in (\ref{e413})
are equal to one half of those from
the interaction terms with matter.
In particular, for the action given in (\ref{schwinger}), 
the use of the equations (\ref{maxwell}), (\ref{amu}) and 
 (\ref{cmu}) allows us to replace its field action 
({\it i.e.}, the part not involving matter current 
explicitly) by
\begin{eqnarray} 
\label{e425}
\!\!\!&&\!\!\!\!\!\!\!\!\int d^4 x\left\{
-{1\over 4}F^{\mu\nu}(\partial_\mu A_\nu-\partial_\nu A_\mu)
-{1\over 4}\,^*\!F^{\mu\nu}
\int \! d^4 x' \! (n\cdot \partial)^{-1}(x,x')
[n_\mu J_{g\nu}(x')
\!-\!n_\nu J_{g\mu}(x')] \right\}\nonumber\\
\!\!\!&&\!\!\!\!\!\!\!\!\sim\! \int\! d^4 x\!\left\{\!
-\!{1\over 4}F^{\mu\nu}\!(\partial_\mu A_\nu\!-
\!\partial_\nu A_\mu)
\!-\!{1\over 2}\!\partial^\mu C^\nu
%\!-\!\partial^\nu C^\mu)
\!\int\! d^4 x'\!  (n\cdot \partial)^{-1}\!(x,x')
[n_\mu J_{g\nu}(x')
\!-\!n_\nu J_{g\mu}(x')]\! \right\}\nonumber\\
\!\!\!&&\!\!\!\!\!\!\!\!\sim\int d^4 x\left\{ 
{1\over 2}J_e^\mu A_\mu +{1\over 2}
J^\mu_g C_\mu(F)\right\},
\end{eqnarray}   
where, on the second line, we have dropped the 
contribution apparently describing string-string
interaction. As analogous reduction holds for
the Higgs field action of (\ref{e413}) also. 
Based on this observation, using the 
solutions (\ref{e417}), (\ref{e420}), (\ref{e421}),
(\ref{e423}) and (\ref{e424}) in the action (\ref{e413})
leads to the effective Lagrangian of the form
\begin{eqnarray}
\int\! dtL\!&=&\! \int dt \biggl\{ 
\!-\!\sum_{n}\! M_n \sqrt{1\!-\!{\dot \bX}^2_n}\!  + 
\!{1 \over 2}\sum_{n,m(\neq n)} 
(q_ng_m\!-\!g_nq_m)(\dot{\bX}_n\!-
\!\dot{\bX}_m)\cdot\omega (\bX_n,\bX_m) \biggr.
\nonumber\\
\!\!\!\!&-&\!{1 \over 8\pi}\sum_{n,m(\neq n)} 
\!(q_nq_m+g_ng_m)\left({1\over |\bX_n-\bX_m|} 
+{1\over 2}\left[ {\partial^2\over \partial t^2} 
|\bx-\bX_m|
\right]_{\bx=\bX_n}
\!\!-{{\dot \bX}_n\cdot{\dot \bX}_m\over 
|\bX_n-\bX_m| }\right)
\nonumber\\
\biggl. \!\!\!\!&+&\!\sum_{n,m(\neq n)}\!\!\!
{(g_s)_n(g_s)_m \over 8\pi} 
\!\!\left({\sqrt{1\!-\!{\dot \bX}^2_n} \sqrt{1\!-
\!{\dot \bX}^2_m}
\over |\bX_n-\bX_m|} 
+{1\over 2}\left[ {\partial^2\over \partial t^2} 
|\bx\!-\!\bX_m|
\right]_{\bx\!=\!\bX_n}\!
\right)\,\,  \biggr\}
\label{e426}
\end{eqnarray}   
with irrelevant self-interaction terms dropped.
Ignoring terms beyond $O(\dot{X}^2)$,
this Lagrangian may then be changed to the form
({\it cf.} Appendix A)
\begin{eqnarray}
\label{e427}
L\!&=&\!-\!\sum_{n} M_n\!+\!
{1\over 2}\sum_{n} M_n {\dot \bX}^2_n \! -\! 
{1\over 16\pi}\sum_{n,m(\neq n)}(g_s)_n(g_s)_m 
{ |\dot\bX_n\!-\!\dot\bX_m|^2 \over |\bX_n\!-\!\bX_m|}
\nonumber\\
\!\!\!\!&+&\!- \!{1 \over 2}\sum_{n,m(\neq n)} 
(q_ng_m-g_nq_m)(\dot{\bX}_n\!-\!\dot{\bX}_m)\cdot\omega 
(\bX_n,\bX_m)
\nonumber\\
\!\!\!\!&-&\!{1\over 16\pi}\!\!\sum_{n,m(\neq n)} \!
((g_s)_n(g_s)_m\!-\!q_nq_m\!-\! g_ng_m)\left\{
{{\dot \bX}_n\cdot{\dot \bX}_m\over |\bX_n\!-\!\bX_m| }\!+\!
{(\bX_n\!-\!\bX_m)\!\cdot\!{\dot \bX}_n (\bX_n\!-\!\bX_m)
\!\cdot\!{\dot \bX}_m
\over |\bX_n-\bX_m|^3 }\right\}
\nonumber\\
\!\!\!\!&+&\!{1\over 8\pi}\!\!\sum_{n,m(\neq n)} {
(g_s)_n(g_s)_m\!-\!q_nq_m\!-\! g_ng_m\over
|\bX_n\!-\!\bX_m|}\ .
\end{eqnarray} 

Some comments are in order as regards the 
slow-motion
effective Lagrangian derived above.
If the given system consists of BPS dyons with the same values
of charges only ({\it i.e.}, 
$q_n=q$, $g_n=g$ and $(g_s)_n=\sqrt{g^2+q^2}$ for all $n$),
all the terms in (\ref{e427}) which are not quadratic in 
velocities
cancel. This is the case in which  {\it static} multi-monopole
solutions are possible, and for some given 
initial velocities the dynamics is governed solely
by the kinetic Lagrangian of the same form as found for
slowly-moving equal-charge
W-particles (see Sec.I). Another case of interest 
follows if we let the magnetic charge of all BPS 
dyons to be equal
({\it i.e.}, 
$g_n=g$ for all $n$) and keep in (\ref{e427}) only terms
which are at most quadratic in velocity or electric charge.
Then, $(g_s)_n\approx g+{q_n^2/2g}$ and the Lagrangian 
(\ref{e427}) reduces to (here, $M=gf$)
\begin{eqnarray}
\label{e428}
L\!&=&\!
{1\over 2}\sum_{n} M ({\dot \bX}^2_n -q_n^2/g^2) \! -\! 
{g^2\over 16\pi}\sum_{n,m(\neq n)} 
{ |\dot\bX_n\!-\!\dot\bX_m|^2 \over |\bX_n\!-\!\bX_m|}
\!+\! 
{1\over 16\pi}\!\!\sum_{n,m(\neq n)} \!
{(q_n-q_m)^2\over |\bX_n\!-\!\bX_m|}
\nonumber\\
\!\!\!\!&+&\!{g \over 2}\sum_{n,m(\neq n)} 
\!(q_n\!-\!q_m)(\dot{\bX}_n\!-\!\dot{\bX}_m)\cdot\omega 
(\bX_n,\bX_m)\ .
\end{eqnarray} 
Using precisely this form, Gibbons and Manton\cite{samols}
showed that one can derive the Lagrangian appropriate
to geodesic motion of $n$ well separated monopoles
on the corresponding multi-monopole moduli space; 
this generalizes the earlier work by Manton\cite{samols}
on the nature of 2-monopole moduli space, where the 
relevant asymptotic metric was known as the self-dual 
Euclidean 
Taub-NUT metric\cite{r23} with a negative mass parameter.
Without repeating this analysis we here only mention that
the electric charge variables $q_n$ in (\ref{e428}) may be 
interpreted as  conserved momenta conjugate
to
the collective coordinates representing $U(1)$ phase
angles of individual monopoles. In conclusion,
our low-energy action (\ref{e413}) predicts
the same physics as the moduli-space geodesic approach 
(for well-separated BPS monopoles of the same 
magnetic charges),
when
the effect of radiation can be ignored. Our action 
(\ref{e413}) can be used to describe low-energy 
processes involving radiation
of the  $A^\mu$ or $\varphi$ explicitly also.

\section{Extension to  More General Gauge Models}
\subsection{Preliminaries}

Up to this point our attention was exclusively
in the context of $SU(2)$ Yang-Mills-Higgs model. We now
want to generalize our discussion to the case of BPS 
dyons appearing 
in a gauge theory with an arbitrary compact 
simple gauge group 
$G$ that is maximally broken to $U(1)^k$ 
($k$ is the rank of $G$). As we shall see, much of the 
structure derived in 
the $G=SU(2)$ model will find direct
generalization to this case.

Using the matrix notations $A_\mu \equiv A^p_\mu T_p$
and $\phi\equiv \phi^p T_p \,\, (p=1, \cdots, 
d={\rm dim} G)$ 
with hermitian generators $T_p$ normalized by
${\rm Tr}(T_p T_q)= \kappa \delta_{pq}$, the Lagrange density 
reads
\begin{equation}
\label{e51}
{\cal L} =-{1\over 4\kappa} {\rm Tr}G^{\mu\nu}G_{\mu\nu}-
{1\over 2\kappa} {\rm Tr}D_\mu\phi D^\mu\phi 
\end{equation}
where
$G^{\mu\nu} \equiv \partial^\mu A^\nu-\partial^\nu A^\mu
-ie[A^\mu, A^\nu]$ and
$(D_\mu \phi)\equiv \partial_\mu\phi-ie[A_{\mu},\phi]$.
As is well-known, generators may be decomposed into
$k$ mutually commuting operators $T_r$, that span 
the Cartan 
subalgebra,
and lowering and raising operators 
$E_{\vec{\alpha}}$ obeying
$[T_r, E_{\vec{\alpha}}]=\alpha_r E_{\vec{\alpha}}$ and 
$[ E_{\vec{\alpha}},  E_{-\vec{\alpha}}]=\sum_{r=1}^k
\alpha_r T_r \,\, (\equiv \vec\alpha \cdot \vec{T})$. 
The nature of the symmetry breaking is determined by the 
asymptotic value of the Higgs field in some fixed 
direction, say, on the positive z-axis. It may be taken 
to lie in the Cartan
subalgebra; this then define a unit vector $\hat{h}$
by
\begin{equation}
\label{e52}
\langle\phi\rangle_v
= \sum_{r=1}^k f \hat{h}_r T_r \equiv f \hat{h} \cdot 
\vec{T},
\end{equation}
where $f$ is some positive number. 
We have a maximal symmetry breaking, {\it i.e.}, 
$G\rightarrow U(1)^k$, if $\hat{h}$ is orthogonal to
none of the root vectors. In the latter case, there 
is a unique set of so-called
simple roots $\vec\beta_r\, (r=1,\cdots, k)$ that satisfies 
the conditions $\hat{h}\cdot\vec\beta_r > 0$ for 
all $r$, and all
other roots can be expressed as linear combinations of
these simple roots with integer 
coefficients all of the same sign.
Only this case will be considered in this paper.

Let us briefly summarize known properties of monopoles/dyons in this
model\cite{r24}.  In the asymptotic region, the magnetic field
$B_i\equiv B^p_i T_p$ must commute with the Higgs field and therefore,
in the spatial direction chosen to define $\langle\phi\rangle_v$, must
assume the form
\begin{equation}
\label{e53}
B_i (\br)
\sim  {x_i\over 4\pi r^3}\vec{g}\cdot\vec{T}\ .
\end{equation}
Topological arguments lead to the quantization condition
\begin{equation}
\label{e54}
\vec{g}= {4\pi\over e}\sum_{r=1}^k n_r \vec{\beta}^*_r \,\,,
\ \ \ \ ( \vec{\beta}^*_r\equiv  \vec{\beta}_r/ 
\vec{\beta}^2_r)
\end{equation}
the nonegative integer $n_r$ being the topologically conserved charges
related to the homotopy class of the Higgs field at spatial infinity.
We may now define the special $U(1)$ electric and magnetic charges
using the asymptotic Higgs field direction as
\begin{eqnarray}
\label{e55}
&&Q_E ={1\over f}\oint_{{r= \infty}} dS_i \,{1\over \kappa}
{\rm Tr}(\phi E_i),\nonumber\\
&&Q_M ={1\over f}\oint_{{r= \infty}} dS_i \,{1\over \kappa}
{\rm Tr}(\phi B_i)\ \ (=\vec{g}\cdot \hat{h}),
\end{eqnarray} 
and similarly the dilaton charge as
\begin{eqnarray}
\label{e56}
Q_S ={1\over f}\oint_{{r= \infty}} dS_i \,{1\over \kappa}
{\rm Tr}(\phi D_i\phi).
\end{eqnarray} 
Then, just as in the $G=SU(2)$ model discussed in Sec. II,
one can show that the mass of a static soliton,
which is always equal to $f Q_S$, satisfies the 
Bogomol'nyi
bound $M\geq f\sqrt{Q_M^2+Q_E^2}$.
Hence, for given values of $Q_E$ and $Q_M$,
one may obtain static solutions
to field equations with the lowest possible energy,
$M=f\sqrt{Q_E^2+Q_M^2}$, by solving again
the Bogomol'nyi equations  which have the same structure
as the corresponding equations of the $SU(2)$
model, {\it viz.}, (\ref{bogomol}). 
Especially, with $Q_E=0$, these lowest energy 
configurations will
have the mass 
\begin{eqnarray}
\label{e57}
M={f}\vec{g}\cdot \hat{h}=\sum_{r=1}^k n_r
\left({4\pi \over e}f
 \hat{h}\cdot\vec{\beta}_r\right).
\end{eqnarray}
On the other hand, Weinberg\cite{r24} showed that 
the dimension 
of the corresponding moduli space is equal to
$4\sum_{r=1}^k n_r$. This suggests that,
in analogy with the $SU(2)$ case, all static solutions
might be viewed as being composed of
a number of {\it fundamental} BPS monopoles,
each with a single unit of topological charge
({\it i.e.}, $n_r=\delta_{rr'}$, for the $r'$-type).

The fundamental static BPS monopole solutions can be 
obtained by
simple embeddings\cite{r25} of the spherically symmetric 
$SU(2)$ solution given in (\ref{monopole}). Note 
that, with each 
root $\vec\alpha$, we can always define an $SU(2)$ 
subalgebra 
with generators
\begin{eqnarray}
\label{e58}
t^1_{(\vec\alpha)}={1\over \sqrt{2\vec\alpha^2}}
(E_{\vec\alpha}+E_{-\vec\alpha}),\  
t^2_{(\vec\alpha)}={i\over \sqrt{2\vec\alpha^2}}
(-E_{\vec\alpha}+E_{-\vec\alpha}), \  t^3_{(\vec\alpha)}
={\vec\alpha\cdot \vec{T}\over \vec\alpha^2}\,.
\end{eqnarray} 
Now, if $\bar{A}^a_i(\br,f)$ and  $\bar{\phi}^a_i(\br,f)$
denotes the static $SU(2)$ BPS monopole
solution corresponding to a Higgs expectation value
$f$ (see (\ref{monopole})), then
\begin{eqnarray}
\label{e59}
&&A_i(\br)=\sum_{i=1}^3 \bar{A}^a_i(\br,f
\hat{h}\cdot 
\vec{\beta}_r) t^a_{(\vec\beta_r)},\nonumber\\
&&\phi_i(\br)=\sum_{i=1}^3 \bar{\phi}^a_i
(\br,f\hat{h}\cdot 
\vec{\beta}_r) t^a_{(\vec\beta_r)}+ f[\hat{h}
-(\hat{h}\cdot 
\vec\beta^*_r)\vec\beta_r]\cdot \vec{T}
\end{eqnarray} 
is the fundamental monopole solution with $\vec{g}=
-{4\pi\over e}
\vec\beta^*_r$ and mass $M_r={4\pi\over e}
f \hat{h}\cdot\vec\beta_r$.
As in the $SU(2)$ case, we can also obtain
dyon solution corresponding to these fundamental 
monopoles by
applying the trick (\ref{dyon}). Here, to push 
the $SU(2)$
analogy further, it will be useful to
write the corresponding asymptotic field strengths 
as\footnote{In a quantized theory, 
the electric charge $q_r$ 
defined by
(\ref{e510}) will be required to be an integer 
multiple of 
$e\vec{\beta}_j^2$}
\begin{eqnarray}
\label{e510}
B_i \sim g_r {x_i\over 4\pi r^3}\hat{r}_a t^a_{(\vec\beta)},
\  E_i \sim q_r {x_i\over 4\pi r^3}\hat{r}_a 
t^a_{(\vec\beta)},
\ D_i\phi \sim -(g_s)_r {x_i\over 4\pi r^3}
\hat{r}_a t^a_{(\vec\beta)},
\end{eqnarray} 
which means, on the positive z-axis ({\it i.e.}, 
the direction chosen to
define $\bar\phi_0$), the behaviors
\begin{eqnarray}
\label{e511}
B_i \sim g_r {x_i\over 4\pi r^3}(\vec\beta^*_r\cdot\vec{T}),
\  E_i \sim q_r {x_i\over 4\pi r^3}(\vec\beta^*_r
\cdot\vec{T}),
\ D_i\phi \sim -(g_s)_r {x_i\over 4\pi r^3}
(\vec\beta^*_r\cdot\vec{T}),
\end{eqnarray} 
For the $r$-type fundamental dyon, we then have 
the values 
$g_r=-{4\pi/e}$, $q_r=g_r\tan\beta$ and 
$(g_s)_r=\sqrt{g_r^2+q^2_r}$;
the mass of this dyon is equal to
$M_r=(g_s)_r f\hat{h}\cdot \vec\beta^*_r$.

\subsection{Low-energy effective theory}

What sort of low-energy 
%interaction 
dynamics for
fundamental BPS dyons follows from the field equations of
the theory?
As in the $SU(2)$ case, one of the most direct 
information on this problem
can be obtained by considering the fundamental dyons in
the presence of some weak asymptotic uniform fields.
Only the asymptotic, gauge or Higgs, field 
strengths
which commute with the Higgs field $\phi$ may be 
allowed here ({i.e.}, the uniform Higgs field 
belonging to
the unbroken $U(1)^k$ subgroup only). We may 
specify the nature of
these applied field strengths by their values
on the $z$-axis where the Higgs field originally there is
$\bar\phi_0=f \hat{h}\cdot \vec{T}$. (This way 
of specifying 
the applied field strengths will have a clear 
physical meaning
if one works in a unitary gauge where the Higgs field is
everywhere aligned in the direction of $\langle \phi \rangle_v$.)
Now the problem is to find the solution
to the field equations, describing the motion
of the $r$-type fundamental dyon in a nonzero asymptotic
field as specified through the conditions
\begin{eqnarray}
\label{e512}
\!\!\!\!&&\!\!\!\! B_i(\br,t) \rightarrow (\vec{B}_0)_i
\cdot\vec{T},
\  E_i (\br,t) \rightarrow (\vec{E}_0)_i
\cdot\vec{T},
\ D_i\phi (\br,t) \rightarrow -(\vec{H}_0)_i
\cdot\vec{T},
\end{eqnarray}
along z-aixs and as $r\rightarrow \infty$.
Here note that $(\vec{B}_0)_i
\cdot\vec{T}\equiv \sum_{r=1}^k (B_0)^r_i T_r$, 
{\it etc.},
and the constant vectors $\vec{B}_0$, $\vec{E}_0$ 
and $\vec{H}_0$
are assumed to be of sufficiently small magnitude.

Remarkably, the desired solution can be given 
using the corresponding solution of $SU(2)$
model, which we discussed in Sec.III. This is the 
generalization of
the embedding procedure described in (\ref{e59}).
Let $\bar{A}^a_\mu (x;f,{\bf B}_0,{\bf E}_0, {\bf H}_0)$
denote the (in general time dependent) $SU(2)$ BPS 
dyon solution 
in the presence of the asymptotic field 
$({\bf B}_0,{\bf E}_0, {\bf H}_0)$.
Then it may directly be verified that
\begin{eqnarray}
\label{e513}
\!\!\!\!\!\!\!\!&&\!\!\!\!\!\!\!\!
A_\mu(x)\!=\!\sum_{i=1}^3 \!\!
\bar{A}^a_i(x;f\hat{h}\!\cdot\! 
\vec{\beta}_r, \vec{{\bf B}}_0\cdot\vec\beta_r,
\vec{{\bf E}}_0\!\cdot\!\vec\beta_r,\vec{{\bf H}}_0
\!\cdot\!\vec\beta_r) 
t^a_{(\vec\beta)}\!+\!x^\lambda[(\vec{G}_0)_{\lambda\mu}
\!-\!
((\vec{G}_0)_{\lambda\mu}\cdot \vec\beta^*_r)
\vec\beta_r]\!\cdot\! \vec{T}\\
\!\!\!\!\!\!\!\!&&\!\!\!\!\!\!\!\!
\phi_i(x)=\sum_{i=1}^3 \bar{\phi}^a_i(
x;f\hat{h}\cdot 
\vec{\beta}_r, \vec{{\bf B}}_0\cdot\vec\beta_r,
\vec{{\bf E}}_0\cdot\vec\beta_r,\vec{{\bf H}}_0
\cdot\vec\beta_r) 
t^a_{(\vec\beta)}+ f[\hat{h}-(\hat{h}\cdot 
\vec\beta^*_r)\vec\beta_r]\cdot \vec{T}\nonumber\\
\!\!\!\!\!\!\!\!&&\!\!\!\!\!\!\!\!\ \ \ \ \ \ \ \ \ - 
x^i[(\vec{H}_0)_i-
((\vec{H}_0)_i\cdot \vec\beta^*_r)\vec\beta_r]\cdot \vec{T}
\label{e514}
\end{eqnarray}
(here, $(\vec{G}_0)_{ij}\equiv\epsilon_{ijk}(\vec{B}_0)_k$ and 
$(\vec{G}_0)^{0i}\equiv (\vec{E}_0)_k$ ) is a solution 
describing
the $r$-type dyon in the nonzero asymptotic
field as specified by (\ref{e512}). Then, based on our
$SU(2)$ solution, we may immediately
conclude that the $r$-type dyon
 in its instantaneous rest frame
should accelerate according to the formula 
(see (\ref{motioneq}))
\begin{equation}
\label{e515}
a_i=-{1\over f \hat{h}\cdot \vec{\beta}_r}[ \cos\beta 
\,(\vec{B}_0)_i
\cdot \vec\beta_r +\sin\beta 
\, (\vec{E}_0)_i
\cdot \vec\beta_r - (\vec{H}_0)_i
\cdot \vec\beta_r]
\end{equation}
which may be rewritten, using the charges defined by 
(\ref{e511}), as
\begin{equation}
\label{e516}
M_r a_i= g_r \vec\beta^*\cdot(\vec{B}_0)_i
+q_r \vec\beta^*\cdot(\vec{E}_0)_i
+(g_s)_r \vec\beta^*\cdot(\vec{H}_0)_i\,\,.
\end{equation}
To find associated long-distance fields 
(including radiation),
recall that, for the $SU(2)$ case, the relevant 
field strengths have 
nonvanishing
components only in the direction of $\hat{r}^a$ 
(or the Higgs field)
and have the amplitude described by ${\bf B}^{\rm em}$, 
${\bf E}^{\rm em}$,
 ${\bf H}$ and $H^0$ through (\ref{radfa})-(\ref{radfd}) 
and (\ref{radfcc}).
 This term implies that, for our solution given by 
(\ref{e513}) and
(\ref{e514}), the corresponding field strengths would 
have the following
large-distance behaviors on the $z$-axis:
\begin{eqnarray}
\label{e517}
{\bf B}_i(\br,t) &\sim& (\vec{B}_0)_i\cdot \vec{T} +
{g_r \vec\beta^*_r\cdot \vec{T}\over 4\pi}\left\{
{(\hat{\bf R}\!-\!
{\bf v}_{\rm ret})_i\over 
(1-\hat{{\bf R}}\cdot{\bf v}_{\rm ret})^3 R^2}+
{[\hat{\bf R}\times(\hat{\bf R}\times \ba)]_i\over 
R}\right\}\nonumber\\
&-&{q_r \vec\beta^*_r\cdot \vec{T}\over 4\pi}\left\{
{(\hat{\bf R}\times {\bf v}_{\rm ret})_i\over R^2}
+{(\hat{\bf R}\times {\bf a})_i\over R}\right\},\\
\label{e518}
{\bf E}_i(\br,t) &\sim& (\vec{E}_0)_i\cdot \vec{T} +
{q_r \vec\beta^*_r\cdot \vec{T}\over 4\pi}\left\{
{(\hat{\bf R}\!-\!
{\bf v}_{\rm ret})_i\over 
(1-\hat{{\bf R}}\cdot{\bf v}_{\rm ret})^3 R^2}+
{[\hat{\bf R}\times(\hat{\bf R}\times \ba)]_i\over 
R}\right\}\nonumber\\
&+&{g_r \vec\beta^*_r\cdot \vec{T}\over 4\pi}\left\{
{(\hat{\bf R}\times {\bf v}_{\rm ret})_i\over R^2}
+{(\hat{\bf R}\times {\bf a})_i\over R}\right\},
\\
\label{e519}
-{D}_i\phi(\br,t) &\sim& (\vec{H}_0)_i\cdot \vec{T} +
{(g_s)_r \vec\beta^*_r\cdot \vec{T}\over 4\pi}
\left\{{(\hat{\bf R}\!-\!
{\bf v}_{\rm ret})_i\over 
(1-\hat{{\bf R}}\cdot{\bf v}_{\rm ret})^3 R^2}+
{(\hat{\bf R}\cdot\ba)\hat{R}_i\over 
R}\right\}\\
\label{e520}
-{D}^0\phi(\br,t) &\sim& 
{(g_s)_r \vec\beta^*_r\cdot \vec{T}\over 4\pi}
\left\{{\hat{\bf R}
\cdot
{\bf v}_{\rm ret}\over 
(1-\hat{{\bf R}}\cdot{\bf v}_{\rm ret})^3 R^2}+
{\hat{\bf R}\cdot\ba\over 
R}\right\}
\end{eqnarray}
Also considering the Lorentz-boosted solution would
change the force law (\ref{e516}) into the corresponding
covariant form ({\it cf}. (\ref{dycolo}))
\begin{eqnarray}
\label{e521}
&&{d\over dt}\left({(M_r-(g_s)_r \vec\beta^*_r\cdot 
X_\mu \vec{H}^\mu){V}_i\over \sqrt{1-{\bf V}^2}}
\right)=
g_r\vec\beta^*_r\cdot [(\vec{B}_0)_i-\epsilon_{ijk} V_j 
(\vec{E}_0)_k]
\nonumber\\
&&\ \ \ \ \ \ \ \ \ \ \ \ \ \ \ \ \ \ \ \ \ \ 
 +
q_r\vec\beta^*_r\cdot [(\vec{E}_0)_i+\epsilon_{ijk} 
V_j (\vec{B}_0)_k]
+(g_s)_r \vec\beta^*_r \cdot(\vec{H}_0)_i\sqrt{1-{\bf V}^2}.
\end{eqnarray}

Without any further analysis, it is clear from the above discussion
that the differences from the $SU(2)$ dyon case are mainly in
prolification of various charges as we have more massless fields.  In
detail we are just seeing that, given the $r$-type fundamental dyon
associated with the root $\vec\beta_r$, it interacts with $k$
different pairs of massless photon and Higgs field (all in a identical
manner as in the $SU(2)$ case), with the strength of its coupling with
the $r'$-th photon or Higgs field set by magnetic $g^{rr'}= g_r
(\vec\beta^*_r)_{r'}$, electric charge $q^{rr'}= q_r
(\vec\beta^*_r)_{r'}$, and dilaton charge $(g_s)^{rr'}= (g_s)_r
(\vec\beta^*_r)_{r'}$. The massless fields here are precisely the ones
one easily identifies by going to the unitary gauge where the Higgs
field is everywhere in the direction of $\bar\phi_0$; the components
lying in the Cartan subalgebra from the gauge and Higgs fields
correspond to nonmassive physical excitations.  The low-energy
effective action may now be written down on the basis of this
observation and the corresponding result in the $SU(2)$ case. The
effective theory would involve a set of position coordinates $\bX_n
\,\,(n=1,\cdots, N)$ for fundamental dyons (the type of which may also
be indicated by the index $n$), $U(1)$ gauge fields $A^{(r)}_\mu
(x)\,\, (r=1, \cdots,k)$ and Higgs fields $\varphi^{(r)}(x)\,\,
(r=1,\cdots,k)$, while the massive vector boson modes are to be
integrated out. We then have the action ({\it cf.} (\ref{e413}))
\begin{eqnarray}
\label{e522}
S_{\rm eff}&=&\int d^4 x \{{1\over 4} F^{(r)\mu\nu}
F_{(r)\mu\nu}
-{1\over 2}F^{(r)\mu\nu}
(\partial_\mu A^{(r)}_{\nu} -\partial_\nu A^{(r)}_{\mu})
-{1\over 2} \partial_\mu \varphi^{(r)} 
\partial^\mu \varphi^{(r)}\}\nonumber\\ 
&+&
\int dt \sum_{n=1}^N \{\,-\!\left(M_n + 
\sum_r(g_s)^{nr} \varphi^{(r)}(\bX_n,t)\right)
\sqrt{1-{\dot \bX}_n^2}\nonumber\\
&\ & \ \ \ -
\sum_r\!q^{nr} [A^{(r)0}(\bX_n,t)\!-\!{\dot\bX}_n\cdot 
{\bf A}^{(r)}
(\bX_n,t)] \nonumber\\
&\ & \ \ \ -\sum_r\!g^{nr} [
C^{(r)0}(\bX_n,t)\!-\!{\dot\bX}_n\cdot {\bf C}^{(r)}
(\bX_n,t)]\,\}
\end{eqnarray} 
with $C^{(r)\mu}$, as  functions of $F^{(r)\mu\nu}$,
defined in the same way
as (\ref{cmusol}). [In (\ref{e522}), the index $n$
in $q^{nr}$, $g^{nr}$ and $(g_s)^{nr}$ is
actually $r'$ if the n-th dyon in question
is of the $r'$-type, {\it viz}., $q^{nr}=q_{r'} 
(\vec\beta^*_{r'})_r$, $g^{nr}=g_{r'} 
(\vec\beta^*_{r'})_r$, {\it etc.}]

The action (\ref{e522}) captures low-energy dynamics of any number of
fundamental BPS dyons (corresponding to various type) and massless
fields in the system. This includes scattering physics involving dyons
and on-shell photons or Higgs particles. Also, for a slowly moving
system of BPS dyons, one may ignore radiation effects and go on to
eliminate all massless fields from this action by using the near-zone
solutions to the respective field equations. This procedure, which
parallel verbatim our discussion in the $SU(2)$ case, leads to the
effective particle Lagrangian which has the same structure as the
$SU(2)$-case Lagrangian (\ref{e426}).  Changes appear just in the
interaction strengths, {\it i.e.}, the second, third, and fourth terms
in the right hand side of (\ref{e426}) now come with the strengths
\begin{eqnarray}
\label{e523}
&&\sum_{r=1}^k (g_s)^{nr}(g_s)^{mr}=(g_s)_n
(g_s)_m\vec\beta^*_n\cdot\vec\beta^*_m ,
\nonumber\\
&& \sum_{r=1}^k (q^{nr}q^{mr}+g^{nr}g^{mr})
\,= (q^{n}q^{m}+g^{n}g^{m})\vec\beta^*_n\cdot
\vec\beta^*_m,\\
&& \sum_{r=1}^k (q^{nr}g^{mr}-g^{nr}q^{mr})
\,= (q^{n}g^{m}-g^{n}q^{m})
\vec\beta^*_n\cdot\vec\beta^*_m,\nonumber
\end{eqnarray}
instead of having the values $(g_s)_n(g_s)_m$,
$(q_{n}q_{m}+g_{n}g_{m})$ and $(q_{n}g_{m}-g_{n}q_{m})$.  Similarly,
when terms beyond $O(\dot{X}^2)$ are ignored, the Lagrangian
(\ref{e427}) is valid for the present case also only if we insert the
multiplicative factor $\vec\beta^*_n\cdot\vec\beta^*_m$ inside the
summation symbol of every term in the right side of (\ref{e427})
except for the first two purely kinematical ones. If one sets
$g_n=g=-{4\pi/ e}$ and further makes the expansion
$(g_s)_n=|g|+{q_n^2/ (2|g|)}$ with this quadratic particle Lagrangian,
one obtain the slow-motion Lagrangian of Lee, Weinberg and
Yi\cite{klee}, which is quadratic not only in velocities but also in
electric charges. Then, as was shown in Ref.\cite{klee}, a simple
Legendre transform may be performed to change the latter into the
Lagrangian appropriate to geodesic motion in the corresponding
multi-monopole moduli space\footnote{It has
recently been shown \cite{klee,r26} that the moduli space metric
obtained by this procedure for distinct fundamental monopoles is in
fact the exact metric over the whole moduli space, {\it i.e.}, for all
values of intermediate distances. This may imply that our effective
action is correct even when two distinct monopoles overlap each
other.}

\section{Discussions}

In this paper an effective field theory approach for BPS dyons and
massless fields has been developed, starting from the analysis of
nonlinear field equations of the Yang-Mills-Higgs system. Our
approach, while being consistent with the moduli-space dynamics of
Manton, can describe low-energy interaction of oppositely-charged BPS
dyon and also process involving radiation of various massless quanta
explicitly. Our discussion was entirely at the classical level, but,
for an appropriately supersymmetrized system, our effective theory
might be generalized to have a quantum significance.  The
electromagnetic duality and (spontaneously broken) scale invariance,
which are manifest in our approach, may play a useful role in such
endeavor.  It would also be desirable to make some contact with the
results of Seiberg and Witten\cite{seiberg}.

There are some interesting related problems which require further
study. To mention few of them,\\ 
(i) Our effective action is correct when all monopoles are separated
in large distance compare with the core size.  If two identical
monopoles overlap, the individual coordinates are not meaningful any
more.  We can describe the low energy dynamics by the geodesic motion
on the Atiyah-Hitchin moduli space.  But radiation, however weak it
may be, should come out from this motion in the moduli space,
including the exchange of the relative charge between two identical
monopoles.  Our point particle approximation does not capture this
physics.  It would be interesting to couple the full moduli space
dynamics to the weak radiation. \\ 
(ii) The present effective field theory approach should be generalized
to the case of full, $N=2$ or $N=4$, super-Yang-Mills system.
Especially the spin effect including the electric and magnetic dipole
moments would appear.   See Ref.\cite{r27} for the corresponding
moduli-space description.\\ 
(iii) For larger gauge groups, we have only considered the cases where
the given simple gauge group  is maximally broken. If a non-Abelian
subgroup remains unbroken, there are fundamental monopoles carrying
non-Abelian magnetic charges and their low-energy dynamics would be
more rich.  (For a recent investigation on this subject, see
Ref.~\cite{r28}.)  Extension of our analysis in this direction would
be most desirable; for instance, one might here consider following the
behavior of the effective theory as one varies the asymptotic Higgs
field from a value giving a purely Abelian symmetry breaking to one
that leaves a non-Abelian subgroup unbroken.

Finally, we should mention the recent work by one of 
the authors and H. Min\cite{r29} where some interesting 
observation was made as regards to the radiation reaction 
 and the finite-size effect in the 
dynamics of the  BPS monopole 
and the duality of these effects against
those of the W-particles.

\acknowledgements

Useful discussion with R. Jackiw, H. Min, P. Yi
and E. Weinberg are acknowledged. 
The work of DB and CL was
supported in part  by 
the Korea Science and Engineering Foundation through
the SRC program of SNU-CTP, and
the Basic Science Research Institute Program
under project No.BSRI-96-2425 and  
No.BSRI-96-2418. The work of KL was supported in part by D.O.E. 
CL also wishes 
to thank the support of the Aspen 
Center for Physics (July, 1997).

\appendix
\setcounter{section}{0}
\setcounter{equation}{0}
\renewcommand{\theequation}{\Alph{section}\arabic{equation}}
\section{Effective Lagrangian for a system of W-particles}

From the low-energy effective action (\ref{lageff}) 
we can derive the 
effective Lagrangian for a system of  slowly moving 
W-particles,
given in (\ref{modlag}), in the following way. The 
field equations for the
massless fields $A^{\mu}(x)$ and $\varphi(x)$ read
\begin{eqnarray} 
\label{amassless}
&&\partial_\nu(\partial^\mu A^\nu-\partial^\nu A^\mu)= 
J^{\mu}(x),
\\
&&\  [J_0(x)=\sum_n q_n\delta^3(\bx-\bX_n(t)),\ \   
{\bf J}(x)=\sum_n q_n\dot\bX_n(t)\delta^3(\bx-\bX_n(t))]
\ \ \ \ \ \ \ \ \ \ \ \ \ \ \  \ \ \phantom{,}\nonumber\\
\label{bmassless}
&&\partial_\nu\partial^\nu \varphi= \sum_n g_s\sqrt{1-
{\dot\bX}_n^2(t)}\delta^3(\bx-\bX_n(t))\equiv J_s(x).
\end{eqnarray}
Assuming slowly varying sources, we may then express the fields 
$A^{\mu}(x)$, $\varphi(x)$ by their usual retarded 
solutions considered 
in the near zone approximation. This gives the 
electromagnetic potential
\begin{eqnarray} 
A^\mu (\bx,t)&=& {1\over 4\pi} \int {J^\mu(\bx', t-
|\bx-\bx'|)\over 
|\bx-\bx'|} d^3 \bx'\nonumber\\
&=&{1\over 4\pi} \int {J^\mu(\bx', t)\over 
|\bx-\bx'|} d^3 \bx'-{1\over 4\pi} {\partial\over\partial t}\left[
\int {J^\mu(\bx', t)} d^3 \bx'\right]\nonumber\\
\label{vpotential}
&+&{1\over 8\pi} {\partial^2\over\partial t^2}\left[
\int  |\bx-\bx'|{J^\mu(\bx', t)} d^3 \bx'\right] + \cdot\cdot\cdot
\end{eqnarray}
and so, for the point sources,
\begin{eqnarray} 
\label{apotential}
A^0 (\bx,t)&=& {1\over 4\pi} \sum_n {q_n\over |\bx-\bX_n(t)|}
+{1\over 8\pi} {\partial^2\over\partial t^2} 
\left( \sum_n q_n |\bx-\bX_n(t)| \right)+ \cdot\cdot\cdot,\\
{\bf A}(\bx,t)&=& {1\over 4\pi} \sum_n {q_n \dot\bX_n\over 
|\bx-\bX_n(t)|}
+ \cdot\cdot\cdot\ \ .
\label{bpotential}
\end{eqnarray}
Similarly, for the Higgs field, we have 
\begin{eqnarray} 
\varphi (\bx,t)&=& -{1\over 4\pi} \int {J_s(\bx', t-
|\bx-\bx'|)\over 
|\bx-\bx'|} d^3 \bx'\nonumber\\
&=& - {1\over 4\pi} \sum_n {g_s\sqrt{1-\dot\bX^2_n}\over 
|\bx-\bX_n(t)|}
+{1\over 4\pi}
{\partial\over\partial t} \left(\sum_n 
{g_s\sqrt{1-\dot\bX^2_n}}\right)\nonumber\\
&-&{1\over 8\pi} {\partial^2\over\partial t^2} 
\left( \sum_n g_s \sqrt{1-\dot\bX^2_n}|\bx-\bX_n(t)| \right)
+ \cdot\cdot\cdot\ \ .
\label{hpotential}
\end{eqnarray}
These expressions may also be obtained 
by considering small-velocity expansion of the known 
Li\'enard-Wiechert-type potentials.

The desired effective Lagrangian for slowly-moving W-particles
is obtained if we eliminate (or integrate  out) the
massless fields $A^{\mu}(x)$ and $\varphi(x)$ from the action 
(\ref{lageff})  by using the above (approximate) solutions 
to the field equations\footnote{This is 
equivalent to more traditional 
approach described,
for instance, in the textbook by Landau and 
Lifshitz\cite{landau}.}.
Here note that, because of (\ref{amassless}) and 
(\ref{bmassless}), the 
contribution from the
massless field action in (\ref{lageff}) can be written 
in the same form as the interaction terms appearing in the 
matter action
$\int dt L_{\rm eff}$. So, to our approximation, the result 
of using 
(\ref{apotential})-(\ref{hpotential}) in the action 
(with irrelevant self-interactions dropped)
is
\begin{eqnarray}
\int dtL\!\!&=&\!\! \int dt \{ 
\!-\!\sum_{n} m_v \sqrt{1\!-\!{\dot \bX}^2_n}  + 
{g_s^2 \over 8\pi}\!\sum_{n,m(\neq n)} 
\!\left({\sqrt{1\!-\!{\dot \bX}^2_n} \sqrt{1\!-\!{\dot \bX}^2_m}
\over |\bX_n-\bX_m|} 
+{1\over 2}\left[ {\partial^2\over \partial t^2} |\bx\!-
\!\bX_m(t)|
\right]_{\bx=\bX_n}\!
\right)
\nonumber\\
&-&{1 \over 8\pi}\sum_{n,m(\neq n)} 
q_nq_m\left({1\over |\bX_n-\bX_m|} 
+{1\over 2}\left[ {\partial^2\over \partial t^2} |\bx-\bX_m(t)|
\right]_{\bx=\bX_n}
\!\!-{{\dot \bX}_n\cdot{\dot \bX}_m\over |\bX_n-\bX_m| }
\right)\}
\label{aaalag}
\end{eqnarray} 
Here notice that
\begin{eqnarray}
\left[ {\partial^2\over \partial t^2} 
|\bx-\bX_m|\right]_{\bx=\bX_n}&=& 
\left[ {\partial\over \partial t} {(\bx-\bX_m(t))\cdot
\dot\bX_m(t))\over|\bx-\bX_m|}\right]_{\bx=\bX_n}\nonumber\\
=\!{{\dot \bX}_n\cdot{\dot \bX}_m\over |\bX_n\!-\!\bX_m| } 
\!\! &-&\!\!
{(\bX_n\!-\!\bX_m)\!\cdot\!{\dot \bX}_n (\bX_n\!-\!\bX_m)
\!\cdot\!
{\dot \bX}_m\over |\bX_n-\bX_m|^3 }
\!-\!{d\over dt} \left[{({\bX}_n\!-\!\bX_m)\!\cdot\!
{\dot\bX}_m\over |\bX_n-\bX_m|}\right]
\label{bbblag}
\end{eqnarray} 
and so, if we ignore terms beyond $O(\dot\bX^2)$ 
and also total time derivative terms from $L$, we obtain the 
Lagrangian of the form
\begin{eqnarray}
\label{ccclag}
L&=&
{1\over 2}\sum_{n} m_v {\dot \bX}^2_n  - 
{g_s^2\over 16\pi}\sum_{n,m(\neq n)} 
{ |\dot\bX_n-\dot\bX_m|^2 \over |\bX_n-\bX_m| }
\nonumber\\
&-&{1\over 16\pi}\!\!\sum_{n,m(\neq n)} \!(g_s^2\!-q_nq_m)
\left\{
{{\dot \bX}_n\cdot{\dot \bX}_m\over |\bX_n\!-\!\bX_m| }\!+\!
{(\bX_n\!-\!\bX_m)\!\cdot\!{\dot \bX}_n (\bX_n\!-\!\bX_m)
\!\cdot\!{\dot \bX}_m
\over |\bX_n-\bX_m|^3 }\right\}
\nonumber\\
&+&{1\over 8\pi}\!\!\sum_{n,m(\neq n)} {g_s^2-q_nq_m\over
|\bX_n\!-\!\bX_m|}\ .
\end{eqnarray} 
As one can easily verify, this can readily be rewritten 
in the form in 
(\ref{modlag}) and (\ref{metric}).

\setcounter{equation}{0}
\renewcommand{\theequation}{\Alph{section}\arabic{equation}}
\section{Derivation of the force law in Lorentz 
boosted  frame}

The system in  (\ref{lag}) is invariant against the 
Lorentz (boost) transformation
\begin{eqnarray} 
\label{atlorentz}
&&t\rightarrow t^*={t+\bv\cdot\br\over \sqrt{1-\bv^2}}, 
\nonumber\\
\label{btlorentz}
&& \br\rightarrow \br^*=\br-(\hbv\cdot\br)\hbv
+{1\over \sss}
((\hbv\cdot\br)\hbv +{\bv t}),
\end{eqnarray}
under which ($A_\mu$, $\phi$) transform as
\begin{eqnarray} 
\label{ctlorentz}
&&A_\mu(x)\rightarrow A^*_\mu(x^*)={dx^\nu\over 
dx^{*\mu}}A_\nu(x),\nonumber\\
\label{dtlorentz}
&& \phi(x)\rightarrow \phi^*(x^*)=\phi(x).
\end{eqnarray}
This of course implies that the fields 
($A^*_\mu(\br, t)$, $\phi^*(\br,t)$) obtained
by the Lorentz boost of an initially given  solution 
($A_\mu(\br,t)$, $\phi(\br,t)$) 
should also satisfy the field equations. 
Here we use this simple observation in order to show 
that the moving dyon 
seen in a different inertial frame obeys the covariant 
equation of motion. 

Let ($A_\mu(\br,t)$, $\phi(\br,t)$) be a dyon solution of the 
field equations (\ref{fieldeq})-(\ref{pfieldeq}), subject to
the constant asymptotic fields ($\bB$,\,$\bE$,\,$\bH$) 
with zero initial (center) velocity. The trajectory of the 
 dyon  will be  governed by the equation of motion 
\begin{equation}
\label{eqomo}
M{d^2\over dt^2}\bX= g {\bf B} +q 
{\bf E}+g_s {\bf H},
\end{equation}
as was shown  in (\ref{dylorentz}). In this reference frame
 the 
asymptotic value of $H^0\,(\equiv -{{\phi^a\over 
|\phi|}(D^0\phi)^a)})$ 
may be taken to be  $O(a^2)$ at most.
%but from  the energy conservation
%of the system,  its leading  term is fixed  to be 
%$\bH\cdot {d\over dt}\bX(t)$.
Then a new solution ($A^*_\mu(\br, t)$, $\phi^*(\br,t)$)  
generated by the Lorentz boost in (\ref{btlorentz}) is associated
with the asymptotic fields ($\bB^*$,\,$\bE^*$)  specified by
\begin{eqnarray} 
&& \bE=(\hbv\cdot\bE^*)\hbv +
{1\over\sss} (\bE^*-(\hbv\cdot\bE^*)\hbv+
\bv\times\bB), \nonumber\\
&& \bB=(\hbv\cdot\bB^*)\hbv +
{1\over\sss} (\bB^*-(\hbv\cdot\bB^*)\hbv-
\bv\times\bE^*) 
\label{eelorentz}
\end{eqnarray}
and  
($H^{*0}$,\,$\bH^*$) by
\begin{eqnarray} 
&&\bH=
{1\over \sss}[(\hbv\cdot\bH^*)\hbv-\bv H^{*0}]+\bH^*-
(\hbv\cdot\bH^*)\hbv, \nonumber\\
\label{h0lorentz}
&& H^0={1\over\sss}[H^{*0}-\bv\cdot\bH^*]\ .
\end{eqnarray}

%The dyon trajectory seen in the original frame, $(t, \bX(t))$,
%and $(t^*, \bX^*(t^*))$ seen from the transformed frame are
Let $X^\mu\equiv(t, \bX(t))$ denotes the 
dyon trajectory seen in the original frame, and 
$X^{*\mu}\equiv(t^*, \bX^*(t^*))$ the trajectory in the 
boosted frame.
Then they should be  
related by [cf. (\ref{btlorentz})]
\begin{eqnarray} 
\label{ttalorentz}
&&t={t^*-\bv\cdot\bX^* \over \sqrt{1-\bv^2}}, \nonumber\\
\label{ttblorentz}
&& \bX=\bX^*-(\hbv\cdot\bX^*)\hbv+{1\over \sss}
((\hbv\cdot\bX^*)\hbv-\bv t^*),
\end{eqnarray}
We may now re-express the each side of (\ref{eqomo}) using
the variables in the boosted frame.
The left hand side is rewritten,  to  $O(a)$, as
\begin{eqnarray} 
M{d^2\over dt^2} \bX(t)\!\!\!&=&\!\!\!
M{dt^*\over dt}{d\over dt^*}\left( 
{dt^*\over dt} {d\over dt^*} \bX(t)  
\right), \nonumber\\
\!\!\!&=&\!\!\!M\left({(\bv\cdot {d {\bf V}^*\over dt^*})\bv 
\over (1-\bv^2)^{3/2}} 
+
{(\hbv\cdot {d {\bf V}^*\over dt^*})\hbv 
\over (1-\bv^2)^{1/2}} 
+{{d {\bf V}^*\over dt^*}-
(\hbv\cdot {d {\bf V}^*\over dt^*})\hbv \over (1-\bv^2)} 
\right),
\label{acca}
\end{eqnarray} 
where ${\bf V}^* = {d\over dt^*}\bX^*$. On the other 
hand, inserting 
(\ref{eelorentz})-(\ref{h0lorentz}) into (\ref{eqomo}), 
we find that 
 the right hand side can be expressed as
\begin{eqnarray} 
g\bB+q\bE+g_s \bH&=&g\{(\hbv\cdot\bB^*)\hbv +
{1\over\sss} (\bB^*-(\hbv\cdot\bB^*)\hbv-
\bv\times\bE^*)\}\nonumber\\
&+&q\{(\hbv\cdot\bE^*)\hbv +
{1\over\sss} (\bE^*-(\hbv\cdot\bE^*)\hbv+
\bv\times\bB)\}\nonumber\\
&+&g_s\{{1\over \sss}
((\hbv\cdot\bH^*)\hbv-\bv H^{*0})+\bH^*-(\hbv\cdot\bH^*)\hbv \}.
\label{llorentz}
\end{eqnarray}
The equation of motion in (\ref{eqomo}) implies that 
the last line  of
(\ref{acca}) should be equal to the 
right hand side of (\ref{llorentz}). Since this is a vector 
equality, the 
components parallel to  $\bv$ on each side should agree, 
so should the 
components perpendicular to $\bv$ on each side.  We multiply each 
perpendicular component by the factor $\sss$, and then 
add the resulting 
perpendicular parts on each side to the parallel parts on
the corresponding side. These operations  lead to the relation,
\begin{eqnarray} 
&&M\left( {(\bv\cdot {d {\bf V}^*\over dt^*})\bv
\over (1-\bv^2)^{3/2}} 
 +
{1 
\over (1-\bv^2)^{1/2}} {d{\bf V}^*\over dt^*} 
 \right)\nonumber\\
&&\ \ \ \ \ \ \ \ \ \ \ \ \ \ \ \ \  =g(\bB^*\!-\!
\bv\times\bE^*)
+q(\bE^*\!+\!
\bv\times\bB)
+g_s\sss\bH^* \!-\! \check{\bf F},
\label{mmorentz}
\end{eqnarray}
where $\check{\bf F}$ is given by
\begin{eqnarray} 
\check{\bf F}&=&g_s{\bv\over\sss}(\bv\cdot\bH^*-H^*_0)
\nonumber\\
&=&g_s{\bv\over\sss}{d\over dt^*}(\bX^*\cdot\bH^*-t^*H^{*0}).
\label{ooorentz}
\end{eqnarray}
Ignoring $O(a^2)$ terms,  we may replace $\bv$ in  
(\ref{mmorentz}) 
by ${\bf V}^*$ since
${\bf V}^*$ is $\bv +O(a)$.
Thus it is now straightforward to find the desired 
covariant equation
\begin{eqnarray} 
{d\over dt^*}\left( {[M -g_sX^*_\mu H^{*\mu}]{\bf V}^*\over 
\sqrt{1-{\bf V}^{*2}}}\right)
=g(\bB^*-
{\bf V}^*\times\bE^*)
+q(\bE^*+
{\bf V}^*\times\bB)
+g_s\bH^* \sqrt{1-{\bf V}^{*2}}.
\label{nnorentz}
\end{eqnarray}
%where $X^{*0}$ denotes $t^*$.

A few comments are in order. First we assumed the acceleration, 
${d{\bf V}\over dt}$ or ${d{\bf V}^*\over dt^*}$, 
to be small as before, and so the above 
covariant equation of the dyon motion 
is of course valid to first order
in the acceleration.  The term $\check{\bf F}$ in 
(\ref{ooorentz}) is 
of second order in the acceleration, but it has been 
included in the above
 covariant equation. 
%that without it the Lorentz covariance of the 
%equation (\ref{nnorentz}) would be spoiled.
%This term, being nonzero from the energy 
%conservation ground, provides  in a sense 
%a minimal way to recover the full covariance.  
The reason comes from the following observation. 
Let us consider the case where the Higgs field
has the  constant asymptotic value $f'\,(\neq f)$. 
If we carried out the same analyses  to find the dyon motion 
with this choice, the mass parameter that enters into
the  dyon equation of motion is $g_s f'$ instead of $g_s f$. 
Hence the change 
in the asymptotic value 
of the Higgs field should be reflected in the mass appearing
in the dyon equation of  motion[{\it cf.} (\ref{weleq})]. 
 This reasoning can be properly taken 
into account if we add the second-order contribution.
%We now note the fact that  
%$-X^*_\mu H^{*\mu}$ is the asymptotic value of the Higgs field
%$|\phi|$. 

\end{document}